# Machine Learning-Assisted Directed Protein Evolution with Combinatorial Libraries

Zachary Wu[a], S. B. Jennifer Kan[a], Russell D. Lewis[b], Bruce J. Wittmann[b], Frances H. Arnold[a,b,1]

[a]Division of Chemistry and Chemical Engineering, California Institute of Technology, Pasadena, CA 91125; and [b]Division of Biology and Bioengineering, California Institute of Technology, Pasadena, CA 91125

[1]To whom correspondence may be addressed. Email: frances@cheme.caltech.edu

**Abstract**
To reduce experimental effort associated with directed protein evolution and to explore the sequence space encoded by mutating multiple positions simultaneously, we incorporate machine learning into the directed evolution workflow. Combinatorial sequence space can be quite expensive to sample experimentally, but machine learning models trained on tested variants provide a fast method for testing sequence space computationally. We validated this approach on a large published empirical fitness landscape for human GB1 binding protein, demonstrating that machine learning-guided directed evolution finds variants with higher fitness than those found by other directed evolution approaches. We then provide an example application in evolving an enzyme to produce each of the two possible product enantiomers (stereodivergence) of a new-to-nature carbene Si–H insertion reaction. The approach predicted libraries enriched in functional enzymes and fixed seven mutations in two rounds of evolution to identify variants for selective catalysis with 93% and 79% *ee*. By greatly increasing throughput with *in silico* modeling, machine learning enhances the quality and diversity of sequence solutions for a protein engineering problem.

**Significance Statement**
Proteins often function poorly when used outside their natural contexts; directed evolution can be used to engineer them to be more efficient in new roles. We propose that the expense of experimentally testing a large number of protein variants can be decreased and the outcome can be improved by incorporating machine learning with directed evolution. Simulations on an empirical fitness landscape demonstrate that the expected performance improvement is greater with this approach. Machine learning-assisted directed evolution from a single parent produced enzyme variants that selectively synthesize the enantiomeric products of a new-to-nature chemical transformation. By exploring multiple mutations simultaneously, machine learning efficiently navigates large regions of sequence space to identify improved proteins and also produces diverse solutions to engineering problems.

**Introduction**
Nature provides countless proteins with untapped potential for technological applications. Rarely optimal for their envisioned human uses, nature's proteins benefit from sequence engineering to enhance performance. Successful engineering is no small feat, however, as protein function is determined by a highly-tuned and dynamic ensemble of states (1). In some cases, engineering to enhance desirable features can be accomplished reliably by directed evolution, in which beneficial mutations are identified and accumulated through an iterative process of mutation and testing hundreds to thousands of variants in each generation (2–4). However, implementing a suitable screen or selection can represent a significant experimental burden.

Given that screening is the bottleneck and most resource-intensive step for the majority of directed evolution efforts, devising ways to screen protein variants *in silico* is highly attractive. Molecular dynamics simulations, which predict dynamic structural changes for protein variants, have been used to predict changes in structure (5) and protein properties caused by mutations (6). However, full simulations are also resource-intensive, requiring hundreds of CPU hours for each variant, a mechanistic understanding of the reaction at hand, and ideally, a reference protein structure. A number of other, less computationally intensive physical models have also been used to identify sequences likely to retain fold and function for further experimental screening (7–9).

An emerging alternative for screening protein function *in silico* is machine learning, which comprises a set of algorithms that make decisions based on data (10). By building models directly from data, machine learning has proven to be a powerful, efficient, and versatile tool for a variety of applications, such as extracting abstract concepts from text and images or beating humans at our most complex games (11, 12). Previous applications of machine learning in protein engineering have identified beneficial mutations (13) and optimal combinations of protein fragments (14) for increased enzyme activity and protein stability, as reviewed recently (15). Here we use machine learning to enhance directed evolution, using combinatorial libraries of mutations to explore sequence space more efficiently than conventional directed evolution with single mutation walks. The size of a mutant library grows exponentially with the number of residues considered for mutation and quickly becomes intractable for experimental screening. However, by leveraging *in silico* models built based on sampling of a combinatorial library, machine learning assists directed evolution to make multiple mutations simultaneously and traverse fitness landscapes more efficiently.

In the machine learning-assisted directed evolution strategy presented here, multiple amino acid residues are randomized in each generation. Sequence-function information sampled from the large combinatorial library is then used to predict a restricted library with an increased probability of containing variants with high fitness. The best-performing variants from the predicted libraries are chosen as the starting points for the next round of evolution, from which further improved variants are identified. We first investigate the benefits of *in silico* screening by machine learning using the dataset collected by Wu and coworkers (16), who studied the effects on antibody binding of mutations at four positions in human GB1 binding protein (theoretical library size $20^4 = 160{,}000$ variants). We then use machine learning-assisted directed evolution to engineer an enzyme for stereodivergent carbon–silicon bond formation, a new-to-nature chemical transformation.

## Results

**Directed evolution and machine learning**
In directed evolution, a library of variants is constructed from parental sequences, screened for desired properties, and the best variants are used to parent the next round of evolution; all other variants are discarded. When machine learning assists directed evolution, sequences and screening data from all the variants can be used to train a panel of models (covering linear, kernel, neural network, and ensemble methods (*SI Appendix*, *Model Training*)). The models with highest accuracy are then used to screen variants in a round of *in silico* evolution, where the models simulate the fitnesses of all possible sequences and rank the sequences by fitness. A restricted library containing the variants with the highest predicted fitnesses is then constructed and screened experimentally.

This work explores the full combinatorial space of mutations at multiple positions. **Fig. 1** illustrates the approach considering a set of four mutated positions. In a conventional directed evolution experiment with sequential single mutations, identifying optimal amino acids for $N$ positions in a set requires $N$ rounds of evolution (**Fig. 1*A***). An alternative directed evolution approach is to randomly sample the combinatorial space, and recombine the best mutations found at each position in a subsequent combinatorial library (**Fig. 1*B***). Machine learning-assisted evolution samples the same combinatorial space with co-mutated positions *in silico*, enabling larger steps through sequence space in each round (**Fig. 1*C***). In this approach, data from a random sample of the combinatorial library, the input library, are used to train machine learning models. These models are used to predict a smaller set of variants, the predicted library, which can be encoded with degenerate codons to test experimentally (17). The best-performing variant is then used as the parent sequence for the next round of evolution with mutations at new positions.

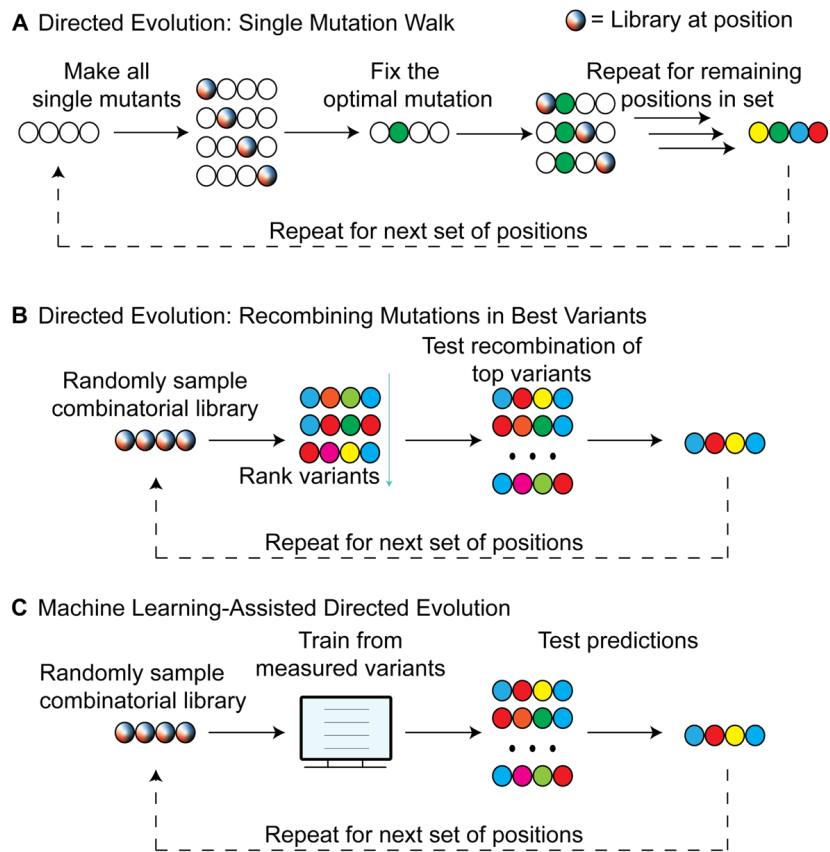

**Figure 1**: (*A*) Directed evolution with single mutations. If limited to single mutations, identifying optimal amino acids for *N* positions requires *N* rounds of evolution. (*B*) Directed evolution by recombining mutations found in best variants from a random combinatorial search. (*C*) Machine learning-assisted directed evolution. Due to increased throughput provided by screening *in silico*, four positions can be explored simultaneously in a single round, enabling a broader search of sequence-function relationships and deeper exploration of epistatic interactions.

**Validation on an Empirical Fitness Landscape**
We first investigated this machine learning-assisted approach on the large empirical fitness landscape of Wu *et al.*, who studied protein G domain B1(GB1) binding to an antibody (16). Specifically, we compare the final fitnesses reached by simulated directed evolution with and without machine learning, based on testing the same number of variants. The empirical landscape used here consists of measurements of 149,361 out of a total $20^4 = 160,000$ variants from NNK/NNS saturation mutagenesis at four positions known to interact epistatically. The fitness of protein GB1 was defined as the enrichment of folded protein bound to the antibody IgG-Fc, measured by coupling mRNA display with next-generation sequencing. The landscape contains a fitness maximum at 8.76, with a fitness value of 1 set for the parent sequence, and 19.7% of variants at a reported value of 0. On this landscape, the simulated single-mutant walk (described below) reached 869 fitness peaks, 533 of which outperformed the wild type sequence and 138 of which had fitness less than 5% of the wild type fitness. For a full description of the epistatic landscape, see the thorough analysis of Wu and coworkers (16).

We first simulated single-mutation evolutionary walks starting from each of the 149,361 variants reported. The algorithm proceeded as follows: In each single-mutation walk, all possible single amino acid mutations were tested at each of the four mutated positions. The best amino acid was then fixed at its observed position, and that position was restricted from further exploration. This process continued iteratively with the remaining positions until an amino acid was fixed at each position. As a greedy search algorithm that always follows the path with strongest improvements in fitness, this single mutation walk has a deterministic solution for each starting variant. Assuming each amino acid occurs with equal frequency and that the library has complete coverage, applying the 3-fold oversampling rule to obtain roughly 95% library coverage (18, 19) results in a total of 570 variants screened (*SI Appendix, Library Coverage*).

Another technique widely used in directed evolution is recombination. For a given set of positions to explore, one method is to randomly sample the combinatorial library and recombine the mutations found at each position in the top $M$ variants. This process is shown in **Fig. 1B**. For $N$ positions, the recombinatorial library then has a maximum of $M^N$ variants, and we selected the top three variants for a maximum recombinatorial library size of 81. An alternative recombination approach is to test all possible single mutants from a given parent sequence and recombine the top three mutations at each position for a fixed recombinatorial library size of 81. However, this alternative recombination does not perform as well on the GB1 data set (*SI Appendix*, **Fig. S1B**). Compared to these recombination strategies, the machine learning-assisted approach has the distinct advantage of providing estimates for the variability at each position (as opposed to taking the top three mutations at each).

To compare the distribution of fitness values of the optimal variants found by the described directed evolution methods, shallow neural networks were trained with 470 randomly-selected input variants, from which 100 predictions were tested, for a total screening burden equivalent to the single-mutation walk. While the number of variants tested was determined by comparison to another method (a single-mutant walk) and the ratio of training variants versus predicted variants was set through experimental convenience (the size of a deep-well plate), from a modeling perspective, these design choices could be improved to increase the expected fitness improvement (*SI Appendix*, **Fig. S1A**). Histograms of the highest fitnesses found by these approaches are shown in **Fig. 2A** and reiterated as empirical cumulative distribution functions in **Fig. 2B**.

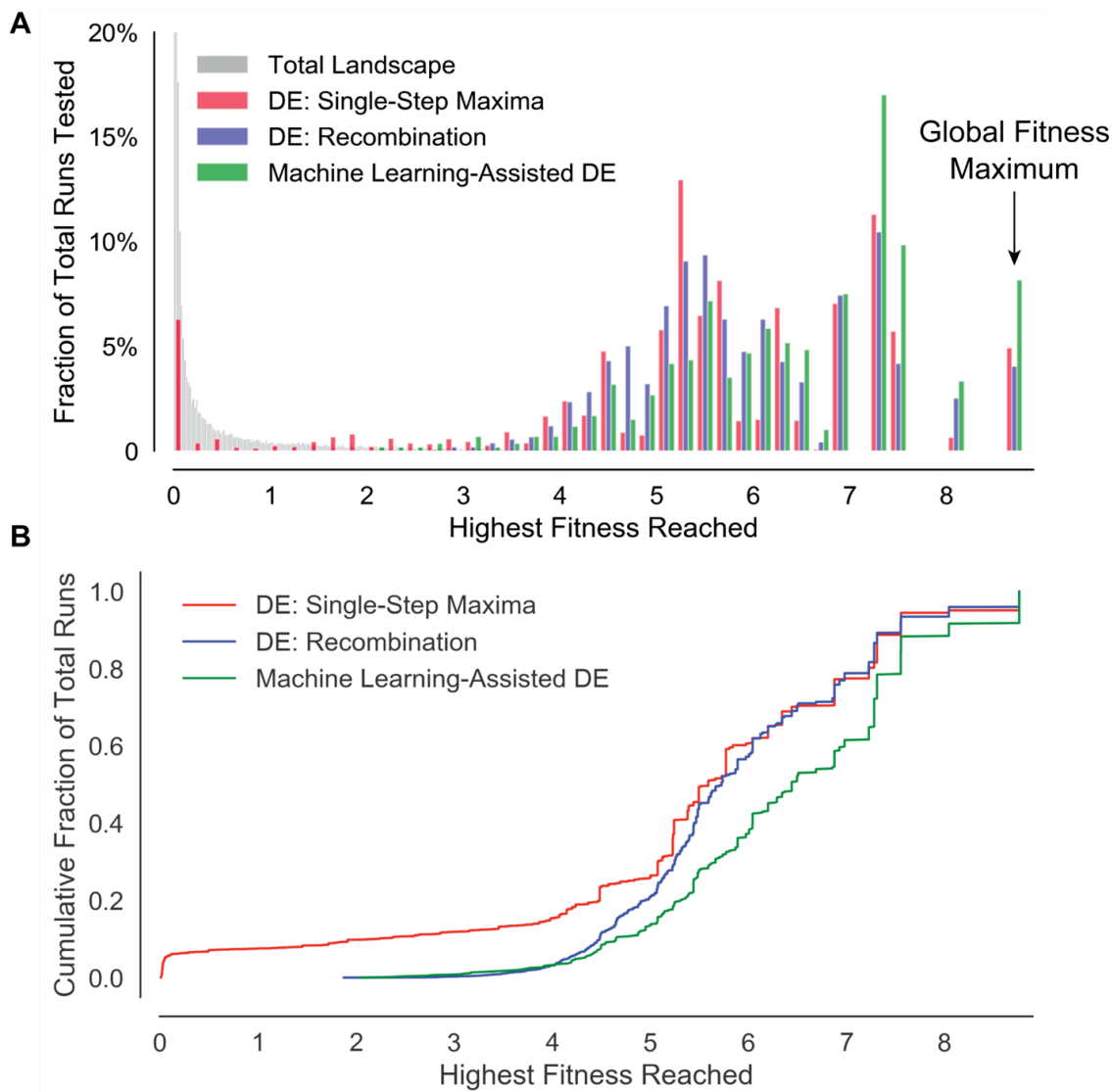

**Figure 2.** (*A*) Highest fitness values found by directed evolution and directed evolution assisted by machine learning. The distribution of fitness peaks found by iterative site-saturation mutagenesis from all labeled variants (149,361 out of $20^4$ possible covering 4 residues) is shown in red. The distribution of fitness peaks found by 10,000 recombination runs with an average of 570 variants tested is shown in blue. The distribution of the highest fitnesses found from 600 runs of the machine learning-assisted approach is shown in green. 570 variants are tested in all approaches. For reference, the distribution of all measured fitness values in the landscape is shown in gray. (*B*) The same evolutionary distributions are shown as empirical cumulative distribution functions, where the ordinate at any specified fitness value is the fraction of evolutionary runs that reach a fitness less than or equal to that specified value. Machine learning-assisted evolution walks are more likely to reach higher fitness levels compared to conventional directed evolution.

As shown in **Fig. 2**, with the same number of variants screened, machine learning-assisted evolution reaches the global optimum fitness value in 8.2% of 600 simulations, compared to 4.9% of all starting sequences reaching the same value through a single-mutant walk and 4.0% of simulated recombination runs. Additionally, on this landscape the machine-learning approach requires about 30% fewer variants to achieve final results similar to the single-mutant walk with this analysis. Perhaps more importantly, a single-mutant walk is much more likely to end at low fitness levels compared to approaches that sample the combinatorial library directly. To this end, the machine learning approach has an expected fitness value of 6.42, compared to 5.41 and 5.93 for the single step walk and recombination, respectively.

Interestingly, the accuracy of the machine learning models as determined on a test set of 1000 random variants not found in the training set can be quite low (Pearson's r = 0.41 with stdev 0.17). However, this level of accuracy as measured by Pearson's r appears to be sufficient to guide evolution. Although perfect accuracy does not seem to be necessary, if the accuracy of the models is so low that predictions are random guesses, this approach cannot be expected to outperform a single mutant walk (*SI Appendix*, **Fig. S1A**). As an algorithm, evolution is focused on identifying optimal variants, and developing a measure of model accuracy biased toward correctly identifying optimal variants will likely improve model selection. This validation experiment gave us confidence that machine learning-assisted directed evolution can find improved protein variants efficiently.

**Application to Evolution of Enantiodivergent Enzyme Activity**
We next used machine learning-assisted directed evolution to engineer an enzyme to produce each of two possible product enantiomers. For this demonstration, we selected the reaction of phenyldimethyl silane with ethyl 2-diazopropanoate (Me-EDA) catalyzed by a putative nitric oxide dioxygenase from *Rhodothermus marinus* (*Rma* NOD), as shown in **Fig. 3**. Carbon–silicon bond formation is a new-to-nature enzyme reaction (20), and *Rma* NOD with mutations Y32K and V97L catalyzes this reaction with 76% *ee* for the *(S)*-enantiomer in whole-cell reactions (*SI Appendix*, **Table 1**).

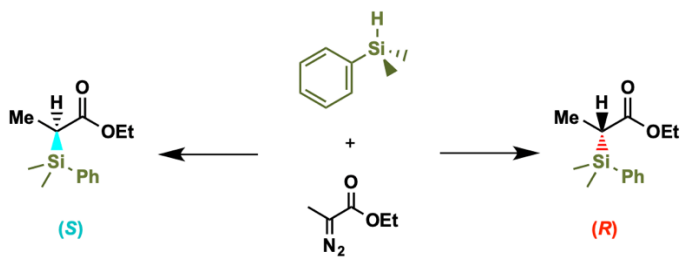

**Figure 3**: Carbon–silicon bond formation catalyzed by heme-containing nitric oxide dioxygenase from *Rhodothermus marinus* to form individual product enantiomers with high selectivity.

Silicon has potential for tuning the pharmaceutical properties of bioactive molecules (21, 22). Because enantiomers of bioactive molecules can have stark differences in their biological effects (23), access to both is important (24). Screening for enantioselectivity, however, typically requires long chiral separations to discover beneficial mutations in a low-throughput screen (25). We thus tested whether machine learning-assisted directed evolution can efficiently generate two catalysts to make each of the product (*S*)- and (*R*)-enantiomers starting from a single parent sequence.

We chose the parent *Rma* NOD (UniProt ID: D0MGT2) (26) enzyme for two reasons. First, *Rma* NOD is native to a hyperthermophile and should be thermostable. Because machine learning-assisted directed evolution makes multiple mutations per iteration, a starting sequence capable of accommodating multiple potentially destabilizing mutations is ideal (27). Second, while we previously engineered a cytochrome *c* (*Rma* cyt *c*) to >99% *ee* for the (*R*)-enantiomer, wild-type *Rma* cyt *c* serendipitously started with 97% *ee* (20). We hypothesized that a parent enzyme with less enantiopreference (76% *ee* for the (*S*)-enantiomer in whole cells) would be a better starting point for engineering enantiodivergent variants.

During evolution for enantioselectivity, we sampled two sets of amino acid positions: Set I contained mutations to residues K32, F46, L56, and L97, and Set II contained mutations to residues P49, R51, and I53 after fixing beneficial mutations identified from Set I. For both sets, we first tested and sequenced an initial set of randomly selected mutants (the input library) to train models. We next tested a restricted set of mutants predicted to have high selectivity (the predicted library). The targeted positions are shown in a structural homology model in **Fig. 4*A***. Set I positions were selected based on proximity to the putative active site, while Set II positions were selected based on their proximity to the putative substrate entry channel.

Machine learning models are more useful when trained with data broadly distributed across input space, even if those data are noisy (28). When designing a training set for machine learning-assisted directed evolution, it is thus important to maximize the input sequence diversity by avoiding disproportionate amino acid representation (e.g. from codon usage). We therefore used NDT codons for the input libraries. NDT libraries encode 12 amino acids having diverse properties with 12 unique codons (18), thus minimizing the probability that an amino acid is overrepresented in the initial training set (29). Notably, the parent amino acid at a site is still considered by the model even if it is not encoded by the NDT codons, as sequence-function data are available for the parent sequence.

The evolution experiment is summarized in **Fig. 4*B***. In the first round, *Rma* NOD Y32K V97L (76 % *ee*) was used as a parent for NDT mutagenesis at the Set I positions. From 124 sequence–function relationships sampled randomly, models were trained to predict a restricted set of selective variants. Specifically, a variety of models covering linear, kernel, shallow neural network, and ensemble methods were tested on each library, from which the optimum models were used to rank every sequence in the theoretical library by its predicted fitness. Under strict limitations in experimental throughput, and with one 96-well plate as the smallest batch size, we settled on two plates of input data for each round of evolution, and one plate of tested predictions. However, increased throughput allows for increased likelihood of reaching the landscape's optimum (*SI Appendix*, **Fig. S1A**). The lower numbers of variants input in **Fig. 4*C*** compared to two full 96-well plates of sequencing reflect failed sequencing reads of these two plates.

From the predicted libraries for both enantiomers, two variants, called VCHV (93 % *ee*) and GSSG (62 % *ee*) for their amino acids at positions 32, 46, 56, and 97, were identified by screening 90 variants for each. VCHV and GSSG were then used as the parent sequences for the second round of mutation at the three positions in Set II. VCHV was the most selective variant in the initial screen, but was less selective in final validation. The approach of experimentally testing a library predicted by models trained on a randomly sampled input library was repeated. From those predicted libraries, we obtained two variants with measured enantioselectivities of 93% and 79% *ee* for the (*S*)- and (*R*)-enantiomers, respectively. These two enantioselective enzymes were achieved after obtaining 445 sequence-function relationships for model training and testing an additional 360 predicted variants, for a total of 805 variants tested experimentally covering 7 positions, as summarized in **Fig. 4C**.

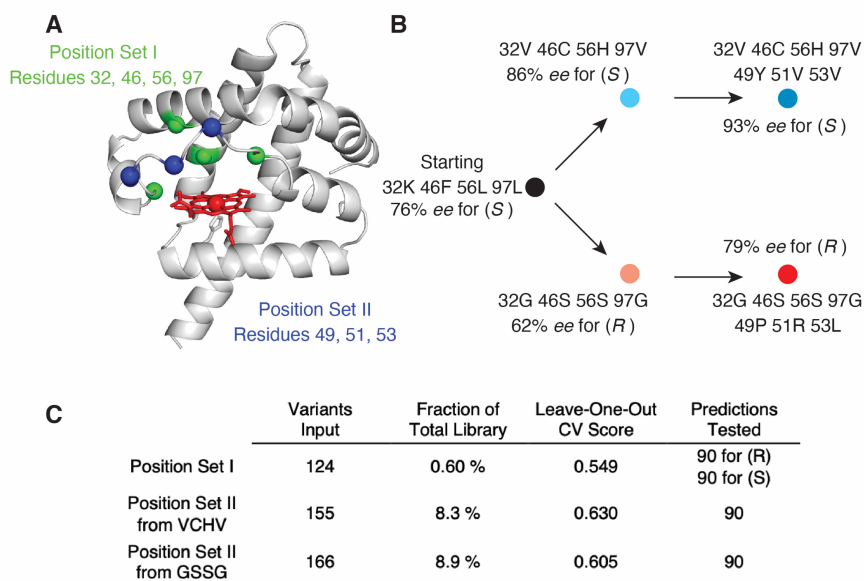

**Figure 4**: (*A*) Structural homology model of *Rma*NOD and positions of mutated residues made by SWISS-MODEL (47). Set I positions 32, 46, 56, and 97 are shown in red, and Set II positions 49, 51, and 53 are shown in blue. (*B*) Evolutionary lineage of the two rounds of evolution. (*C*) Summary statistics for each round, including the number of sequences obtained to train each model, the fraction of the total library represented in the input variants, each model's Leave-One-Out Pearson correlation, and the number of predicted sequences tested.

## Machine Learning Identifies Diverse Improved Sequences

Comparison on the empirical protein GB1 dataset showed that machine learning-assisted directed evolution is more likely than directed evolution alone to identify improved variants. Yet another benefit of this approach is the ability to identify a diverse set of sequences for accomplishing a specific task. Having diverse solutions is attractive as some of those variants may satisfy other design requirements, such as increased total activity, altered substrate tolerance, specific amino acid handles for further protein modification, or sequence diversity for intellectual property considerations (30). By enabling exploration of the combinatorial space, machine learning-assisted directed evolution is able to identify multiple solutions for each engineering objective.

**Table 1** and **Table 2** summarize the most selective variants in the input and predicted libraries for position Sets I and II. The input library for Set I is the same for both product enantiomers. The parent sequences for Set II, VCHV and GSSG, are highlighted in cyan and red, respectively, in the tables. The improvement in total activity measured in whole cells compared to the starting variant (32K, 46F, 56L, 97L) obtained after two rounds of machine learning-assisted directed evolution is also shown in **Table 2.** Although evolved for enantioselectivity, the variants have increased levels of (cellular) activity. Negative controls with cells expressing non-heme proteins yield a racemic mixture of product enantiomers, due to a low level of nonselective background activity from free heme or heme proteins. Increasing the cellular activity of the *Rma*NOD protein can overcome this background activity and appears in the screen as improved selectivity if the protein is selective. Thus, enhanced activity is one path to higher selectivity. The two variants most selective for the (*S*)-enantiomer differ by less than 1% in *ee*. However, the 49P 51V 53I variant from VCHV has higher total activity under screening conditions. By providing multiple solutions in a combinatorial space for a single design criterion, machine learning is able to identify variants with other beneficial properties.

The solutions identified by this approach can also be non-obvious. For example, the three most (*S*)-selective variants in the initial input for Position Set I are YNL**L**, CSV**L**, and CVH**V**. The three most selective sequences from the restricted, predicted library are VGV**L**, CFN**L**, and VCH**V**. If only considering the last residue in bold, the predicted library can be sampled from the top variants in the input library. However, for each of the other three positions, there is at least one mutation that is not present in the top three input sequences.

**Table 1.** Summary of the most (S)- and (R)-selective variants in the input and predicted libraries in Set I (K32, F46, L56, L97). The parent sequences used for Set II for (S)- and (R)-selectivity are shown in cyan and red, respectively.

| Set I: Residues 32, 46, 56, and 97 | | | | | | | | | |
|---|---|---|---|---|---|---|---|---|---|
| Input Variants | | | | | Predicted Variants | | | | |
| Residue | | | | Selectivity | Residue | | | | Selectivity |
| 32 | 46 | 56 | 97 | % ee | 32 | 46 | 56 | 97 | % ee |
| Y | N | L | L | 84 % (S) | V | G | V | L | 90 % (S) |
| C | S | V | L | 83 % (S) | C | F | N | L | 90 % (S) |
| C | V | H | V | 82 % (S) | V | C | H | V | 86 % (S) |
| C | R | S | G | 56 % (R) | G | S | S | G | 62 % (R) |
| I | S | C | G | 55 % (R) | G | F | L | R | 24 % (R) |
| N | V | R | I | 47 % (R) | H | C | S | R | 17 % (R) |

**Table 2.** Summary of the most (S)- and (R)-selective variants in the input and predicted libraries in Position Set II (P49, R51, I53). Mutations that improve selectivity for the (S)-enantiomer appear in the background of [32V, 46C, 56H, 97V (VCHV)] and for the (R)-enantiomer are in [32G, 46S, 56S, 97G (GSSG)]. Activity increase over the starting variant, 32K, 46F, 56L, 97L (KFLL), is shown for the final variants. The parent sequences used for evolving for (S)- and (R)-selectivity are highlighted in cyan and red, respectively.

| | Set II: Residues 49, 51, 53 | | | | | | | | |
|---|---|---|---|---|---|---|---|---|---|
| | Input Variants | | | | Predicted Variants | | | | Cellular Activity |
| | Residue | | | Selectivity | Residue | | | Selectivity | increase over KFLL |
| | 49 | 51 | 53 | % ee | 49 | 51 | 53 | % ee | |
| Evolved | P | R | I | 86 % (S) | Y | V | V | 93 % (S) | 2.8-fold |
| from | Y | V | F | 86 % (S) | P | V | I | 93 % (S) | 3.2-fold |
| VCHV | N | D | V | 75 % (S) | P | V | V | 92 % (S) | 3.1-fold |
| Evolved | P | R | I | 62 % (R) | P | R | L | 79 % (R) | 2.2-fold |
| from | Y | F | F | 57 % (R) | P | G | L | 75 % (R) | 2.1-fold |
| GSSG | C | V | N | 52 % (R) | P | F | F | 70 % (R) | 2.2-fold |

**Machine Learning Predicts Regions of Sequence Space Enriched in Function**

While the machine learning-assisted approach is more likely to reach sequences with higher fitness, as demonstrated in simulations using the human GB1 dataset, there may well be instances where other evolution strategies serendipitously discover variants with higher fitness more quickly. Therefore, since the purpose of library creation is to increase likelihood of success, we caution against focusing solely on examples of individual variants with higher fitness and propose an alternative analysis.

Sequence-fitness landscapes are typically represented with fitness values on the vertical axis, dependent on some ordering of the corresponding protein sequences. Representing this high-dimensional space, even when it is explored with single mutations, is complicated and requires sequencing each variant (31). However, in functional protein space, the engineer is primarily concerned with fitness. Therefore, an alternative representation of a library is a 1-dimensional distribution of fitness values sampled at random for each encoded library. In other words, the sequences are disregarded for visualization, and the library is represented by the distribution of its fitness values. Each subplot in **Fig. 5** shows both the input and predicted (output) library as kernel density estimates in each round of evolution for R- and S-selectivity as fitness. This representation shows the main benefit of incorporating machine learning into directed evolution, which is the ability to focus expensive experiments on regions of sequence space enriched in desired variants.

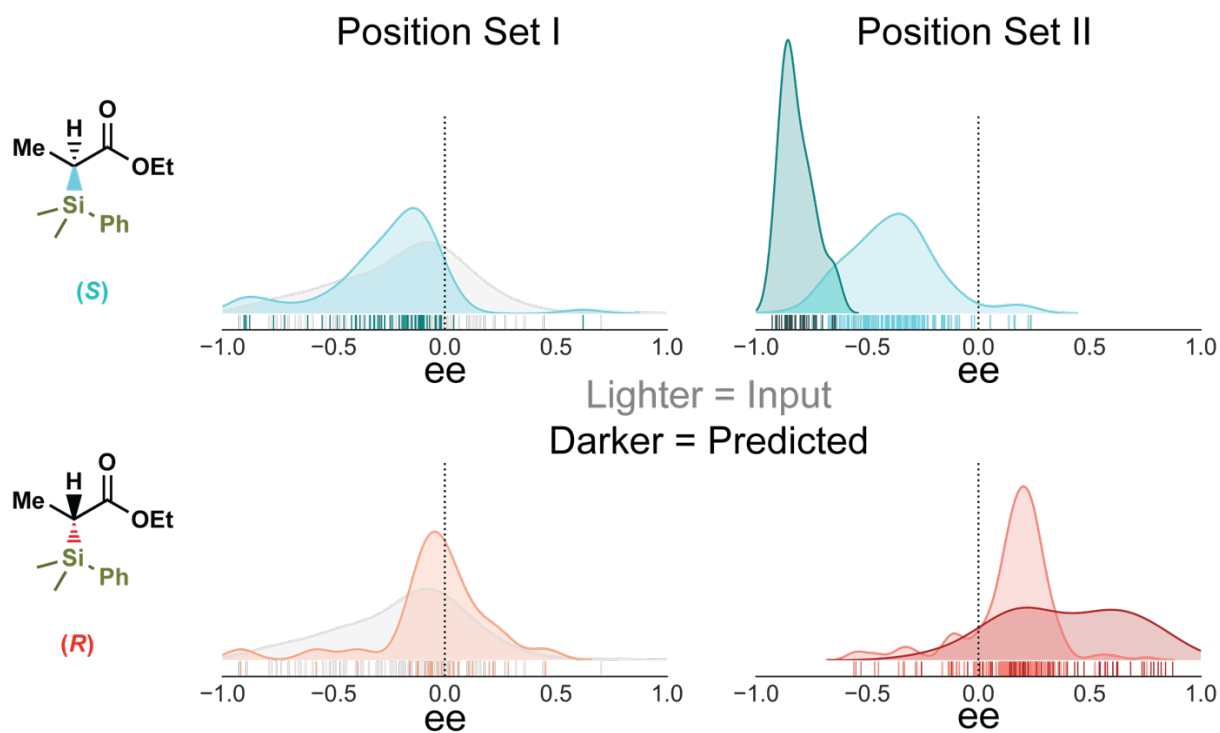

**Figure 5**: A library's fitness values can be visualized as a 1-dimensional distribution, in this case as kernel density estimates over corresponding rug plots. This figure shows subplots for each library illustrating the changes between input (lighter) and predicted (darker) libraries for the (*S*)- (cyan) and (*R*)-enantiomers (red). The initial input library for Set I is shown in gray. The predicted (darker) libraries for each round are shifted toward the right and left of the distributions for the (*S*)- and (*R*)-enantiomers, respectively. For reference, dotted lines are shown for no enantiopreference (0% *ee*).

A few things are immediately clear with this visualization. First, the distribution of random mutations made in the input libraries is shifted toward the parent in **Fig. 5**, as has been shown previously (32). In other words, random mutations made from an (*R*)-selective variant are more likely to be (*R*)-selective. More importantly, the machine-learning algorithm is able to focus its predictions on areas of sequence space that are enriched in high fitness, as can be seen in the shift in distribution from input to predicted libraries. Specifically, 90 predicted variants were tested for predicted library sizes of 864 for the (*S*)-enantiomer and 630 for the (*R*)-enantiomer in Position Set I. In Position Set II, the predicted library sizes were much smaller, at 192 and 90 variants for the (*S*)- and (*R*)- enantiomer, respectively. Ninety variants were tested for these predicted libraries, which were sequenced for redundancy (due to the smaller theoretical library size) to yield 47 and 39 unique variants. Thus, machine learning optimized directed evolution by sampling regions of sequence space dense in functionality. Notably, the machine learning algorithms appear to have more pronounced benefits in Position Set II, likely due to the smaller number of positions explored and larger number of sequence-function relationships obtained.

**Discussion**

We have shown that machine learning can be used to quickly screen a full recombination library *in silico* using sequence-fitness relationships randomly sampled from the library. The predictions for the most-fit sequences are useful when incorporated into directed evolution. By sampling large regions of sequence space *in silico* to reduce *in vitro* screening efforts, we rapidly evolved a single parent enzyme to generate variants that selectively form both product enantiomers of a new-to-nature C–Si bond-forming reaction. Rather than relying on identifying beneficial single mutations as other methods such as ProSAR do (13), we modeled epistatic interactions at the mutated positions by sampling the combinatorial sequence space directly and incorporating models with nonlinear interactions.

Machine learning increases effective throughput by providing an efficient computational method for estimating desired properties of all possible proteins in a large library. Thus we can take larger steps through sequence space by identifying combinations of beneficial mutations, circumventing the need for indirect paths (16) or alterations of the nature of selection (31), and potentially avoiding negative epistatic effects resulting from the accumulation of large numbers of mutations (33) that require reversion later in the evolution (34). This gives rise to novel protein sequences that would not be found just by recombining the best amino acids at each position. Allowing simultaneous incorporation of multiple mutations accelerates directed evolution by navigating different regions of the fitness landscape concurrently and avoiding scenarios where the search for beneficial mutations ends in low-fitness regions of sequence space.

Importantly, machine learning-assisted directed evolution also results in solutions that appear quite distinct. For example, proline is conserved at residue 49 in two of the most (*S*)-selective variants. Proline is considered unique for the conformational rigidity it confers, and at first may seem structurally important, if not critical for protein function. However, tyrosine and arginine are also tolerated at position 49 with less than 1% loss in enantioselectivity. This suggests that there are diverse solutions in protein space for specific properties, as has also recently been shown in protein design (8). Computational models make abstractions to efficiently model physical processes, and the level of abstraction must be tailored to the task, such as protein structure prediction (35). While predictive accuracy could be improved by more computationally-expensive simulations or by collecting more data for machine learning, improved variants can already be identified by sampling from a space predicted to be dense in higher fitness variants. Nevertheless, full datasets collected with higher throughput methods such as deep mutational scanning (36) serve as valuable test beds for validating the latest machine-learning algorithms for both regression (37, 38) and design (39) that require more data.

An evolution strategy similar in spirit to that described here was recently applied to the evolution of GFP fluorescence (40). However, the implementations are quite different. Saito and coworkers used Gaussian processes to rank sequences based on their probability of improvement, or the probability that a variant outperforms those in the training set (39). We take a different approach of identifying the optimal variants, focusing efforts in the area of sequence space with highest fitness. Additionally, because it is difficult to know *a priori* which models will be most accurate for describing a particular landscape, we tested multiple types of models, from linear to ensemble models, to predict the optimal sequences. Modeling the effects of previously-identified point mutations has also recently been studied for evolution of enantioselectivity of an enzyme (41). This study and others focused on increasing the accuracy of protein modeling by developing other physical descriptors (42, 43) or embedded representations (44) suggest that machine learning will assist directed evolution beyond the baseline implementation employed here.

By providing an efficient estimate for desired properties, machine learning models are able to leverage the information from limited experimental resources to model proteins, without the need for a detailed understanding of how they function. Machine learning-assisted directed evolution with combinatorial libraries provides a tool for understanding the protein sequence-function relationship and for rapidly engineering useful proteins. Protein engineers have been sentenced to long treks through sequence space in the search for improved fitness. Machine learning can help guide us to the highest peaks.

**Materials and Methods**
**Approach Validation on an Empirical Fitness Landscape**
Fitness values were provided for 149,361 out of 160,000 total possible sequences covering 4 positions in human protein GB1, where fitness was defined as the enrichment of folded protein bound to IgG-Fc antibody as measured by coupling mRNA display with next-generation sequencing (16). We only use measured sequences and did not incorporate imputed values of variants that were not measured directly. Three directed evolution approaches were simulated on this landscape: 1) a single mutation walk, 2) simulated recombination, and 3) directed evolution with machine learning. For 1), the algorithm proceeds as follows: i) From a starting sequence, every possible single mutation ($19N$ variants for $N$ positions) is made and evaluated. ii) The best single mutation is fixed in the reference sequence, and the position it was found in is locked from further editing. iii) Steps i and ii are repeated until every position has been tested, for a total of 4 rounds to cover 4 positions. 2) Simulated recombination proceeds by selecting 527 random variants and recombining the mutations found in the top three variants, for an average of 570 variants tested over 10,000 simulations. 3) Directed evolution with machine learning proceeds as follows: i) 470 randomly-selected sequences in the combinatorial space are used to train shallow neural networks with randomized hyperparameter search from 4-fold cross-validation based on Pearson's r. Errors are then calculated based on 1000 randomly selected variants that were not present in the training set. ii) The optimal model is used to predict the top 100 sequences, or roughly the screening capacity of a plate. iii) The highest true fitness value in this predicted set of 100 sequences and the training set of 470 is the maximum fitness value found. This process was repeated with different numbers of random sequences in step i to simulate lower model accuracies, the results of which can be seen in **Supp Fig. 1A**. In **Fig. 2**, 100 variants was used as the size of the predicted library test for its similarity to the screening capacity of a 96-well plate. With 570 total variants (*SI Appendix, Library Coverage)* this leaves 470 variants for the input library in step I for an equal screening burden, assuming 95% coverage of 19 mutations from wild type at each position.

**Library Cloning, Expression, and Characterization of *Rma* NOD**
The gene encoding *Rma* NOD was obtained as a gBlock and cloned into pET22b(+) (Novagen catalog number 69744). Standard PCR amplification and Gibson assembly were used for libraries with degenerate codons specified by SwiftLib(17). Encoded versus sequenced codon distributions are shown in **Supp Fig. 2**. Expression was performed in 96 deep-well plates in 1mL HyperBroth (AthenaES) using *Escherichia coli* BL21 *E. cloni* EXPRESS (Lucigen) with 100 μg/mL ampicillin from a 20-fold dilution of overnight culture. Expression cultures were induced after 2.5 hours of outgrowth with 0.5 mM and heme production was enhanced with supplementation of 1 mM 5-aminolevulinic acid.

The relative product activity was measured using 10 mM Me-EDA and 10 mM PhMe$_2$SiH with whole cells resuspended in 400 uL nitrogen-free M9-N buffer, pH 7.4 (47.7 mM Na$_2$HPO$_4$, 22.0 mM KH$_2$PO$_4$, 8.6 mM NaCl, 2.0 mM MgSO$_4$, and 0.1 mM CaCl$_2$). Reactions were incubated anaerobically at room temperatures for 6 hr, before extraction into 600 uL cyclohexane. Enantiomeric excess was measured by running the organic solution on a JACSO 2000 series supercritical fluid chromatography (SFC) system with a Chiralcel OD-H (4.6 mm x 25 cm) chiral column (95% CO$_2$, 5% isopropanol, 3 minutes).

## *Rma* NOD Model Training and Prediction Testing

Screening information was paired with protein sequence obtained from rolling circle amplification followed by sequencing by MCLab. The sequence-function pairs, available on ProtaBank (45), were used to train a panel of models with default hyperparameters in the scikit-learn Python package (46), including K-nearest neighbors, linear (including Automatic Relevance Detection, Bayesian Ridge, Elastic Net, Lasso LARS, and Ridge), decision trees, random forests (including AdaBoost, Bagging, and Gradient Boosting), and multilayer perceptrons. The top 3 model types were selected, and gridsearch cross-validation was used to identify the optimal hyperparameters. The top 3 hyperparameter sets for the top 3 model types were used to identify the top 1000 sequences in each predicted library. Degenerate codons encoding amino acids occurring with highest frequencies in every model at each position were identified by Swiftlib (17), and 90 random variants were tested *in vitro*. This random sampling differs from that in the empirical fitness landscape, where all sequences have been enumerated and can be easily tested. While sampling randomly means we may not have tested the optimal sequence as identified in trained models, we are able to generate fitness distributions as in **Fig. 6B** to describe this space.

**Acknowledgements**:
The authors would like to thank Yisong Yue for initial guidance and Scott Virgil at the Caltech Center for Catalysis and Chemical Synthesis (3CS) for providing critical instrument support. Additionally, we would like to thank Kevin Yang, Anders Knight, Oliver Brandenburg, and Ruijie Kelly Zhang for helpful discussions. This work is supported by the National Science Foundation GRF2017227007 (to Z.W.), the Rothenberg Innovation Initiative (RI2) Program (S.B.J.K. and F.H.A.), and the Jacobs Institute for Molecular Engineering for Medicine at Caltech (S.B.J.K. and F.H.A.).

Supporting Information for
**Machine Learning-Assisted Protein Evolution with Combinatorial Libraries**

Zachary Wu[a], S. B. Jennifer Kan[a], Russell D. Lewis[b], Bruce J. Wittmann[b], Frances H. Arnold[a,b,1]

[a]Division of Chemistry and Chemical Engineering, California Institute of Technology, Pasadena, CA 91125; and [b]Division of Biology and Bioengineering, California Institute of Technology, Pasadena, CA 91125

[1]To whom correspondence may be addressed. Email: frances@cheme.caltech.edu

*This section is particularly useful for those who would like to implement this method.

I.   General procedures

   (A) Plasmid construction

All variants described in this study were cloned and expressed using the pET22(b)+ vector (MilliporeSigma, St. Louis, MO). The gene encoding wild-type *Rhodothermus marinus* putative nitric oxide dioxygenase (*Rma* NOD, UniProt ID (1): D0MGT2_RHOM4) was obtained as a single gBlock (Integrated DNA Technologies, Coralville, IA), codon-optimized, and cloned using Gibson assembly (2) into pet22(b)+ with a 6xHisTag appended at the C-terminus. This plasmid was transformed into *E. cloni* ® EXPRESS BL21(DE3) cells (Lucigen, Middleton, WI).

DNA coding sequence of *Rma* NOD 32K 97L with a C-terminal 6xHisTag:

ATGGCGCCGACCCTGTCGGAACAGACCCGTCAGTTGGTACGTGCGTCTGTGCCTGCA
CTGCAGAAACACTCAGTCGCTATTAGCGCCACGATGTATCGGCTGCTTTTCGAACGG
TATCCCGAAACGCGGAGCTTATTTGAACTTCCTGAGAGACAGATACACAAGCTTGCG
TCGGCCCTGTTGGCCTACGCCCGTAGTATCGACAACCCATCGGCGTTACAGGCGGCC
ATCCGCCGCATGGTGCTTTCCCACGCACGCGCAGGAGTGCAGGCCGTCCATTATCCG
CTGGTTTGGGAATGTTTGAGAGACGCTATAAAAGAAGTCCTGGGCCCGGATGCCAC
CGAGACCCTTCTGCAGGCGTGGAAGGAAGCCTATGATTTTTAGCTCATTTACTGTC
TACCAAGGAAGCGCAAGTCTACGCTGTGTTAGCTGAACTCGAGCACCACCACCACC
ACCACTGA

Amino acid sequence of *Rma* NOD 32K 97L with a C-terminal 6xHisTag

MAPTLSEQTRQLVRASVPALQKHSVAISATMYRLLFERYPETRSLFELPERQIHKLASAL
LAYARSIDNPSALQAAIRRMVLSHARAGVQAVHYPLVWECLRDAIKEVLGPDATETLL
QAWKEAYDFLAHLLSTKEAQVYAVLAELEHHHHHH

   (B) Protein expression

Single colonies from Luria Broth (LB)-ampicillin (100 μg/mL) agar plates were picked using sterile toothpicks and grown in 600 μL LB-ampicillin in 2 mL 96 deep-well plates at 37°C, 250 rpm, 80% humidity overnight (12-18 hours). Multi-channel pipettes were used to transfer 50 μL of overnight culture into 4 deep-well plates containing 1 mL Hyperbroth (HB, AthenaES) each. Four replicates of each well position were made to minimize variability in cell culture and maximize accuracy for downstream modeling. The expression plate were incubated at 37°C, 250 rpm, 80% humidity for 2.5 hours. The plates were then chilled on ice for 30 minutes and induced with 0.5 mM isopropyl β-D-1-thiogalactopyranoside and supplemented with 1 mM 5-aminolevulinic acid to increase heme production. The plate was incubated at 22°C, 220 rpm overnight. The plate was then centrifuged at 3000g for 10 minutes at 4°C. Each individual well was resuspended in 100 μL M9-N buffer (pH 7.4, 47.7 mM $Na_2HPO_4$, 22.0 mM $KH_2PO_4$, 8.6 mM NaCl, 2.0 mM $MgSO_4$, and 0.1 mM $CaCl_2$). The four replicates were combined for 400 μL total in M9-N buffer.

### (C) Biocatalytic reaction and assay

In an anaerobic chamber, 10 μL of 400 mM PhMe$_2$SiH (in acetonitrile) and 10 μL of 400 mM ethyl-2-diazopropanoate (Me-EDA, in acetonitrile) were added to 380 μL whole-cells resuspended in M9-N buffer. The final concentrations in each well were 10 mM Me-EDA and 10 mM PhMe$_2$SiH in each 400 μL reaction mix. The reaction plate was covered with a foil cover (USA scientific) and shaken at 1000 rpm for 4 hours. 600 μL of cyclohexane was added with a multi-channel pipette to each well to quench the reaction and extract the reaction products, which have been previously characterized (3). Plates were centrifuged to remove cells (3000g, 10 minutes) and enantiomeric excess was measured by running the organic solution on a JACSO 2000 series supercritical fluid chromatography (SFC) system with a Chiralcel OD-H (4.6 mm x 25 cm) chiral column (95% CO$_2$, 5% isopropanol, 3 minutes). Final variants in main text **Table 1a** and **Table 1b** were expressed and tested in biological triplicate (in addition to the previous protocol of combining 4 replicates). Automatic integration was performed in ChemStation.

### (D) Model training

Machine-learning models were trained with sequencing information from MCLAB Inc and enantiomeric data obtained by SFC. To model the data, the following regressors from the superlative scikit-learn package (4) were used: K-nearest neighbors, linear (including Automatic Relevance Detection, Bayesian Ridge, Elastic Net, Lasso LARS, and Ridge), decision trees, random forests (including AdaBoost, Bagging, and Gradient Boosting), and multilayer perceptrons, as it is difficult to know *a priori* which model will best fit the landscape. For example, if the selected positions are truly non-interactive, we can expect much of the landscape's variance to be explained by a linear model. However, for more epistatic landscapes, we must account for this nonlinearity. Therefore, many different model classes were tested, all of which can be run (with hyperparameter optimization) on a personal MacBook Pro in less than one day. The 3 model types with highest Pearson correlation from a Leave-One-Out cross validation (LOO CV) with default hyperparameters were selected for gridsearch hyperparameter optimization. From this gridsearch, the 3 sets of hyperparameters with highest LOO CV Pearson correlation were selected, for a total of 9 models in order to capture different characteristics of the landscape with relatively low accuracy models. The models were retrained on the full dataset and used for predicting a restricted library, discussed in the section below.

### (E) Model predictions

Directly synthesizing the DNA encoding the top variants is quite expensive. Therefore, we interpret our models' predictions by the frequency of each amino acid's occurrence in a top fraction (the top 1000) of the library, which we are able to encode efficiently with degenerate codon libraries. An example is shown in **supplementary Table 3**. For cloning purposes, at this point the sequence information predicted from the models is lost, as each position is considered independently to reduce DNA synthesis and subcloning costs. Additionally, we elected to include all 20 amino acids in the predictions even though less than 20 were encoded in the input libraries, to provide an estimate for when the models may be predicting high fitness based on mutations at other positions. A full description of this step is provided with the accompanying **supplementary Table 3**.

### (F) Experimental validation of predictions

The top amino acids at each position are encoded by degenerate codons identified by SwiftLib (5). All 9 models are considered when choosing amino acids to encode, in case some models are capturing different characteristics of the sequence-function relationship. While the optimal combinations of amino acids identified by the model are retained in this library, there may be non-optimal combinations that result from this procedure. However, we have developed this method to balance these experimental costs with being able to access the restricted libraries. The degenerate codons used to encode the predicted libraries are shown in **supplementary Figure 3.** The predicted libraries were tested in the same manner as above.

## Table 1: Starting activity

Although WT has slightly lower enantioselectivity (and thus may reach both enantiomers more easily), we started with a previously-engineered variant, Y32K V97L, for its significantly higher activity, which we hypothesized would make data collection more reproducible. Activity and selectivity are reported in biological triplicate.

| Variant | (S)-enantiomer Area (mAU*s) | (R)-enantiomer Area (mAU*s) | Enantiomeric Excess |
|---|---|---|---|
| Rma NOD WT | 1350 ± 90 | 340 ± 30 | 59 % |
| Rma NOD Y32K V97L | 2710 ± 50 | 370 ± 20 | 76% |

## Table 2: Modeling statistics

The accuracy for the 9 models of each Set, as well as the average values of the 9 models, is shown for the data obtained by screening the predicted libraries. The corresponding predicted versus measured values can be found in **supplementary Figure 3**.

### Table 2a: Test errors for Set I from predicted (R)- and (S)- libraries

| Position Set I | Kendall tau | MAE | Pearson r |
|---|---|---|---|
| Model_0 | 0.32518 | 30.01188 | 0.480996 |
| Model_1 | 0.32518 | 30.0119 | 0.480996 |
| Model_2 | 0.32518 | 30.01191 | 0.480996 |
| Model_3 | 0.382935 | 22.62033 | 0.471391 |
| Model_4 | 0.452322 | 27.23348 | 0.548192 |
| Model_5 | 0.347679 | 32.51843 | 0.428702 |
| Model_6 | 0.362329 | 30.58998 | 0.514802 |
| Model_7 | 0.365468 | 30.38801 | 0.516186 |
| Model_8 | 0.404083 | 17.22301 | 0.508073 |
| Average | 0.390582 | 26.07764 | 0.517561 |

**Table 2b: Test errors for Set II from predicted (*R*)- library**

| Position Set II (*R*)- | Kendall tau | MAE | Pearson r |
|---|---|---|---|
| Model_0 | 0.608254 | 0.44737 | 0.822691 |
| Model_1 | 0.608254 | 0.447369 | 0.822691 |
| Model_2 | 0.595273 | 0.446588 | 0.818083 |
| Model_3 | 0.28028 | 0.223076 | 0.385406 |
| Model_4 | 0.626798 | 0.194871 | 0.827613 |
| Model_5 | 0.626798 | 0.196244 | 0.821946 |
| Model_6 | 0.598982 | 0.410763 | 0.823711 |
| Model_7 | 0.598982 | 0.410763 | 0.823711 |
| Model_8 | 0.598982 | 0.410763 | 0.823711 |
| Average | 0.610108 | 0.285607 | 0.820453 |

**Table 2c: Test errors for Set II from predicted (*S*)- library**

| Position Set II (*S*)- | Kendall tau | MAE | Pearson r |
|---|---|---|---|
| Model_0 | -0.03104 | 0.230254 | -0.04925 |
| Model_1 | 0.020243 | 0.209426 | 0.072554 |
| Model_2 | 0.041835 | 0.216491 | 0.098214 |
| Model_3 | 0.22807 | 0.109102 | 0.342213 |
| Model_4 | 0.265857 | 0.123463 | 0.363029 |
| Model_5 | 0.063428 | 0.139915 | 0.118149 |
| Model_6 | 0.086663 | 0.215941 | 0.144677 |
| Model_7 | 0.086663 | 0.215941 | 0.144677 |
| Model_8 | 0.086663 | 0.215941 | 0.144677 |
| Average | 0.060729 | 0.1903 | 0.108533 |

**Table 3: Sample prediction frequency table – position Set II (*S*)-**
A sample predicted frequency table from position Set II for the (*S*)-enantiomer is provided below. At this step, the exact combinations predicted by the machine learning models are lost, and instead interpreted as frequencies at individual amino acid positions to encode with degenerate codons. The alternative (ordering and cloning the top sequences individually) can be quite expensive, but tractable if screening costs significantly outweigh DNA synthesis costs. All 20 canonical amino acids are enumerated, although each library does not contain all 20 in the input library. Amino acids that are not present in the input library, but predicted to be high-functioning, can be used as indicators for when the frequencies may be relying on amino acids at other positions to make predictions. In other words, they serve as cut-offs above which the amino acids should be considered.

| AA1 | Freq1 | AA2 | Freq2 | AA3 | Freq3 |
|---|---|---|---|---|---|
| Y | 153 | V | 203 | I | 302 |
| N | 115 | F | 123 | V | 288 |
| R | 91 | I | 80 | L | 93 |
| G | 81 | Y | 54 | S | 73 |
| Q | 50 | Q | 48 | F | 51 |
| S | 49 | W | 46 | M | 27 |
| T | 47 | H | 45 | Q | 21 |
| W | 47 | M | 43 | K | 19 |
| K | 47 | L | 43 | T | 19 |
| E | 46 | A | 43 | P | 19 |
| A | 45 | E | 40 | A | 18 |
| H | 37 | K | 40 | W | 17 |
| V | 37 | T | 37 | E | 13 |
| M | 36 | R | 36 | G | 9 |
| C | 35 | P | 36 | H | 9 |
| D | 31 | S | 32 | Y | 7 |
| F | 21 | C | 27 | N | 6 |
| L | 12 | N | 10 | C | 5 |
| P | 11 | G | 8 | D | 2 |
| I | 9 | D | 6 | R | 2 |

For example, the first amino acids that were not present in the input dataset for each of the three positions are Q, Q, M as NDT encodes: {N, S, I, H, R, L, D, G, V, Y, C, F}. The amino acids occurring significantly more frequently in the top 1000 sequences are then [Y, N, R, G], [V, F, I], and [I, V, L, S, F]. This process is repeated for the top 3 models determined by default hyperparameters, and then 3 hyperparameter sets are used for each optimal, for a total of 9 models (in an attempt to capture different portions of the landscape with inaccurate models). In the described approach, this step can be tuned to consider more or less sequences (such as the top 20% or top 1%) depending on the protein engineer's discretion considering the following: screening throughput, sequencing cost, cloning capabilities, desired fitness improvement, model accuracy, theoretical size of the predicted library, ease of encoding with codon degeneracy, and an interpretation of the landscape (how many variants are expected to be near the fitness peak). As DNA synthesis costs continue to fall, the ideal is to be able to sample the top sequences directly (as was simulated in our study with data from a full recombination library of 4 positions) without resorting to this approach to interpret the models' predictions with degenerate codons.

**Supplementary Table 4: Relative activity compared to starting sequence**

The relative activity compared to KFLL is shown for the top 3 variants from each input and predicted round.

### Set I: Residues 32, 46, 56, and 97

| | Input Variants | | | | | Input Variants | | | | |
|---|---|---|---|---|---|---|---|---|---|---|
| | **Residue** | | | | **Activity compared to KFLL** | **Residue** | | | | **Activity compared to KFLL** |
| | 32 | 46 | 56 | 97 | | 32 | 46 | 56 | 97 | |
| | Y | N | L | L | 2.0 | V | G | V | L | 1.9 |
| (S)- | C | S | V | L | 1.7 | C | F | N | L | 2.1 |
| | C | V | H | V | 2.4 | V | C | H | V | 2.5 |
| | C | R | S | G | 2.2 | G | S | S | G | 2.7 |
| (R)- | I | S | C | G | 2.0 | G | F | L | R | 1.1 |
| | N | V | R | I | 2.3 | H | C | S | R | 0.9 |

### Set II: Residues 49, 51, and 53

| | | Input Variants | | | | | Predicted Variants | | | | | |
|---|---|---|---|---|---|---|---|---|---|---|---|---|
| | | **Residue** | | | **Enantioselectivity** | | **Activity compared to KFLL** | **Residue** | | | **Enantioselectivity** | | **Activity compared to KFLL** |
| | | 49 | 51 | 53 | % S-isomer | % R-isomer | | 49 | 51 | 53 | % S-isomer | % R-isomer | |
| (S)-selective From VCHV | | P | R | I | **93** | 7 | 2.5 | Y | V | V | **97** | 3 | 2.8-fold |
| | | Y | V | F | **93** | 7 | 1.5 | P | V | I | **96** | 4 | 3.2-fold |
| | | N | D | V | **87** | 13 | 0.8 | P | V | V | **96** | 4 | 3.1-fold |
| (R)-selective From GSSG | | P | R | I | 19 | **81** | 2.7 | P | R | L | 11 | **89** | 2.2-fold |
| | | Y | F | F | 22 | **78** | 0.8 | P | G | L | 13 | **87** | 2.1-fold |
| | | C | V | N | 24 | **76** | 0.6 | P | F | F | 15 | **85** | 2.2-fold |

**Supplementary Figure 1A:** Highest fitnesses found with less accurate models

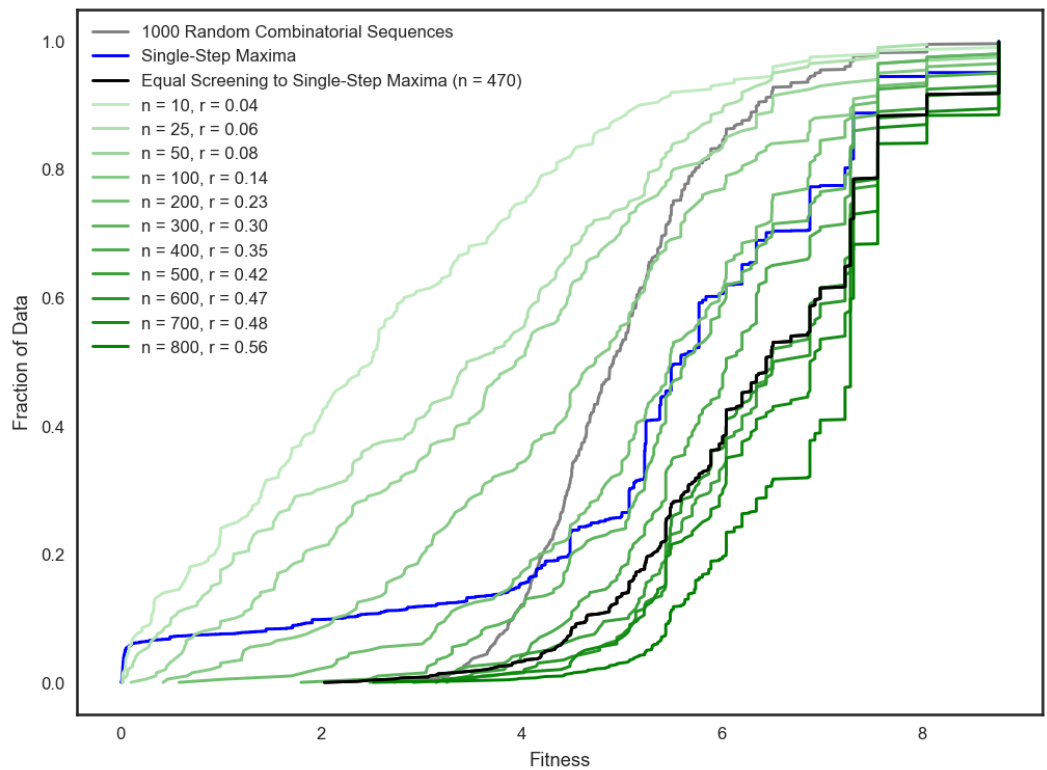

**Supplementary Figure 1B:** Highest fitnesses found with other DE approaches

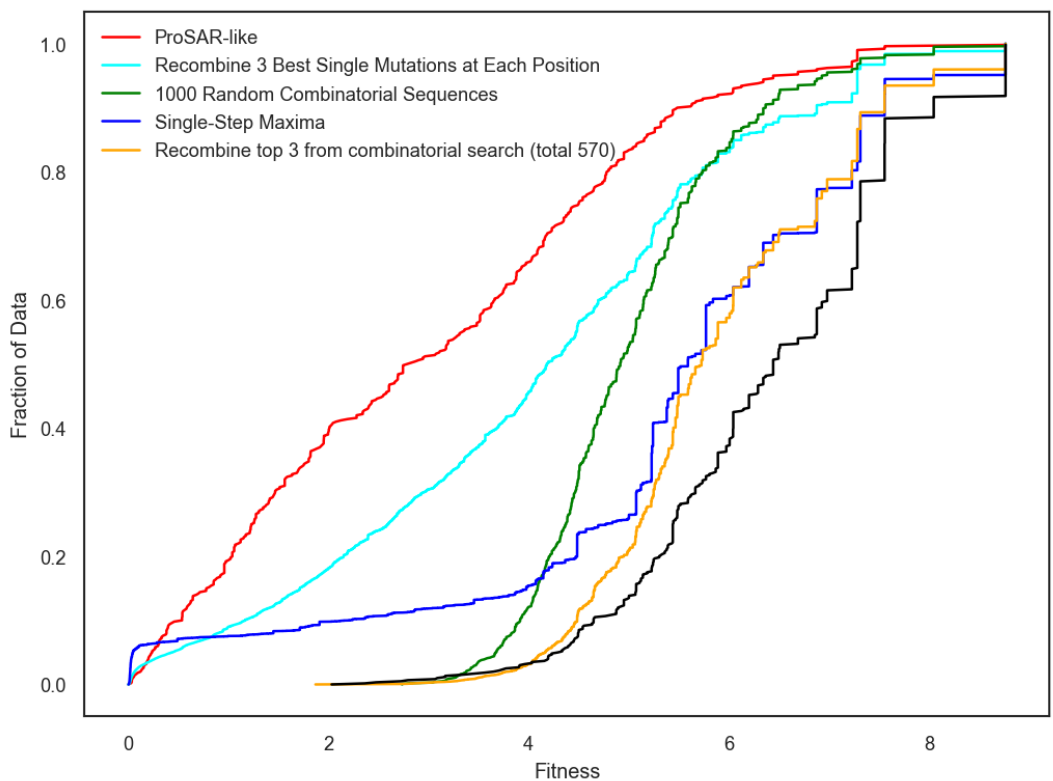

Empirical Cumulative Distribution Functions (eCDFs) are shown for increasing amount of data input in **Supp. Fig. 1A**, and for the various evolutionary methods established in the main text in **Supp. Fig. 1B**. In **Supp. Fig. 1A**, two hundred simulated evolutions are tested for each of the machine learning-assisted methods, with the exception of N=470, where 600 are tested. Lines shifted toward the right are more likely to identify sequences with higher fitness. The cumulative fraction is shown on the ordinate axis, and fitness value on the abscissa. The highest fitness value from the top 100 sequences (roughly the smallest batch size, as screening is typically done in 96 well plates) from each model trained with N sequences is shown to demonstrate the effect of increased training data. Therefore, the total screening burden for each line is N + 100. With 570 sequences measured (in black), the machine learning-assisted evolution approach reaches the global optimum fitness value in 8.4% of simulations, compared to 4.9% of all starting sequences (in blue). The machine learning-assisted evolution approach only requires between 300 and 400 total tested sequences to perform similarly to directed evolution (570 sequences). Therefore, the directed evolution approach requires about 42% more variants tested to achieve similar results on this landscape. However, perhaps a more important metric is the expected fitness value obtained by each method, summarized below.

|  | Expected Fitness Reached (equivalent screening) | Fraction of Runs that reach the Maximum |
|---|---|---|
| ProSAR | 3.00 | 0.20% |
| Recombining 3 Best Single Mutations at Each Position | 4.07 | 1.18% |
| 1000 Random Combinatorial Sequences | 5.04 | 0.40% |
| Single Step Mutation Walk | 5.41 | 4.91% |
| DE+ML (300 total sequences) | 5.46 | 3.5% |
| DE+ML (400 total sequences) | 5.74 | 2.0% |
| Testing random sequences, and recombining the top 3 | 5.93 | 4.03% |
| **DE+ML (570 total sequences)** | **6.42** | **8.17%** |

Other controls are included in **Supp. Fig. 1B for** random combinatorial sequences, from which the highest fitness from 1000 random samples is provided (in gray), two different methods of recombination (in cyan and gold), and a ProSAR-like algorithm (in red). In cyan, recombination from the top 3 single mutants at each position from a reference parent are shown. The top 3 mutants from a random combinatorial search of all positions is shown in gold (with an average of 570 sequences searched).

Our implementation of ProSAR is based on the Partial Least Squares (PLS) algorithm for a linear model for point mutations established by Fox and coworkers (6, 7). Specifically, the PLS implementation by scikit-learn is trained with data from 569 random sequences (optimized over the number of components kept). From the PLS decomposition, the coefficients for the linear contribution from each mutation is determined, and the most positive mutation at each position is kept. We call this approach "ProSAR-like", as the exact implementation of ProSAR can be fairly subjective (see Supporting Information (Detailed description of a round): *Improving catalytic function by ProSAR-driven enzyme evolution* by Fox *et al.* (7)).

The low performance of ProSAR on this landscape is worth discussing. ProSAR was developed to analyze previously-identified mutations at different positions, such that each position typically only has one (maybe two) mutations to consider. A base model with linear contributions at these positions supported their evolution. However, in our recombination landscape of a small number of positions with known epistasis (nonlinear effects), this approach should not be expected to find optimal solutions (and does not outperform other methods tested).

**Supplementary Figure 2**: **Predicted vs measured values for all libraries**
The predicted versus measured values for sequence-verified variants in the predicted libraries are shown for each library. Figure 2A contains predicted values for position Set I. Figure 2B contains predicted values for position Set II from GSSG, and Figure 2C for Set II from VCHV. A linear regression is shown for these values.

**Figure 2A**: Predicted vs measured values for *ee* from Set I

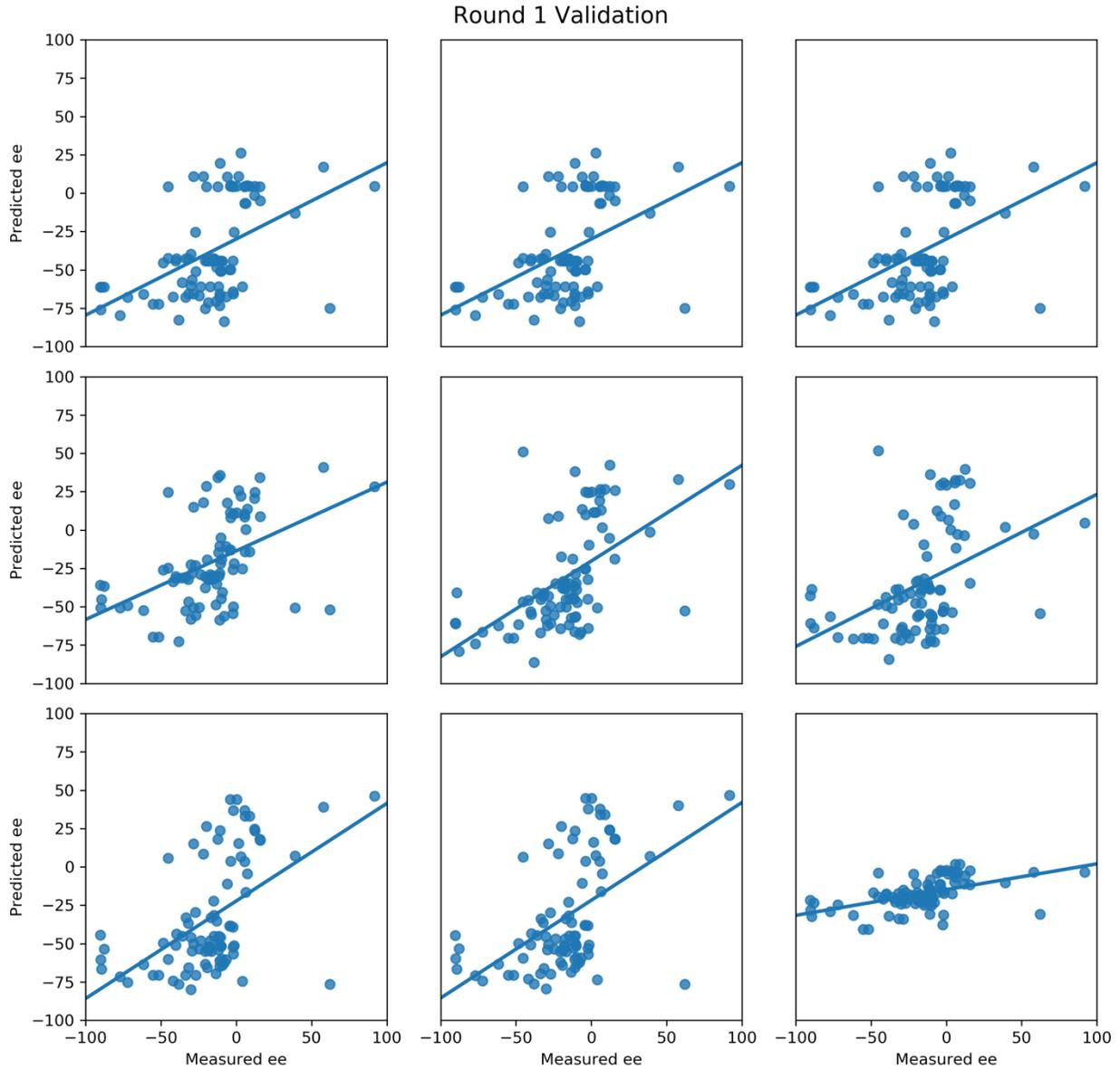

**Figure 2B**: Predicted vs measured values for *ee* from position Set II from GSSG

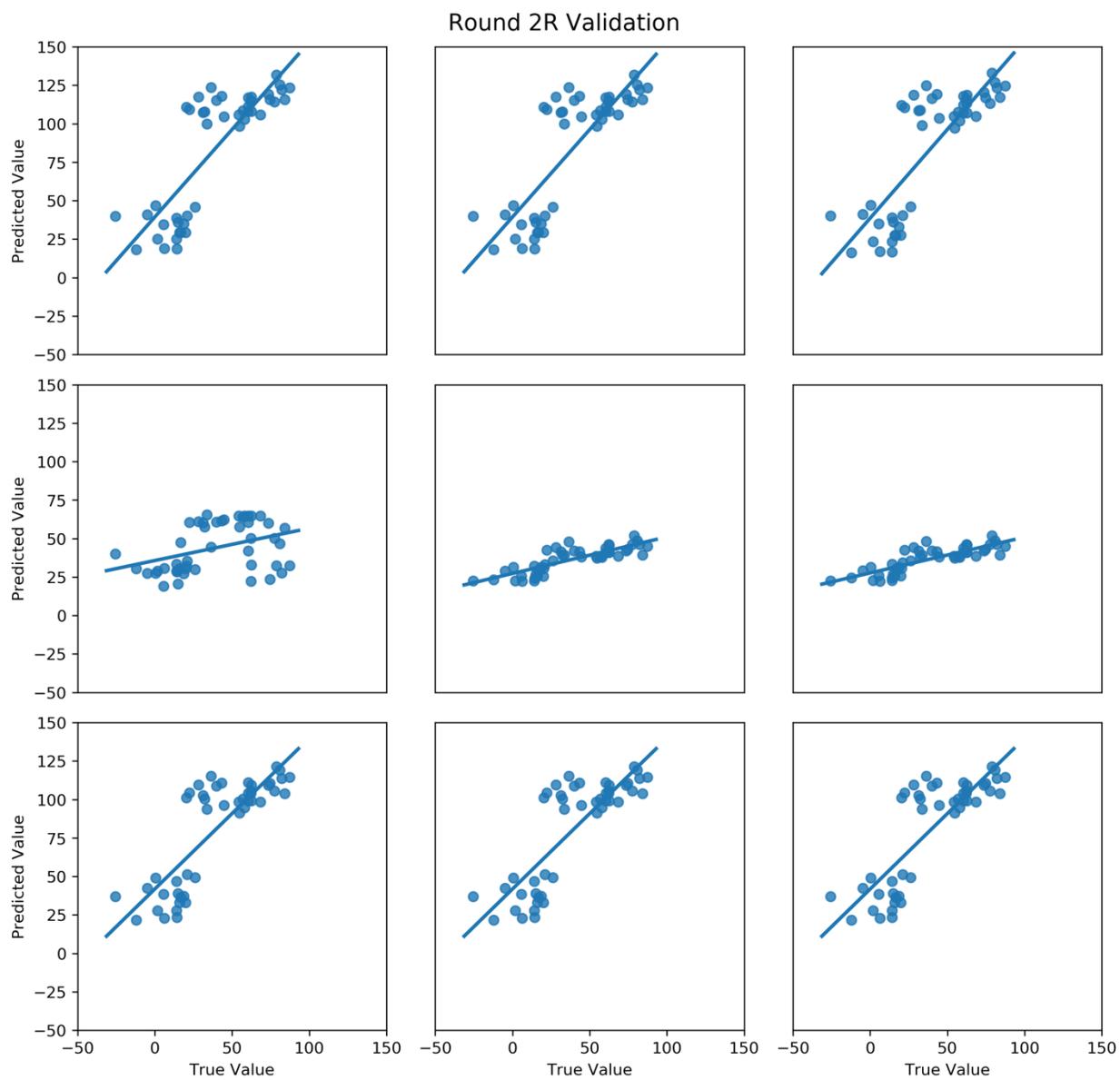

**Figure 2C**: Predicted vs Measured Values for *ee* from Position Set II from VCHV

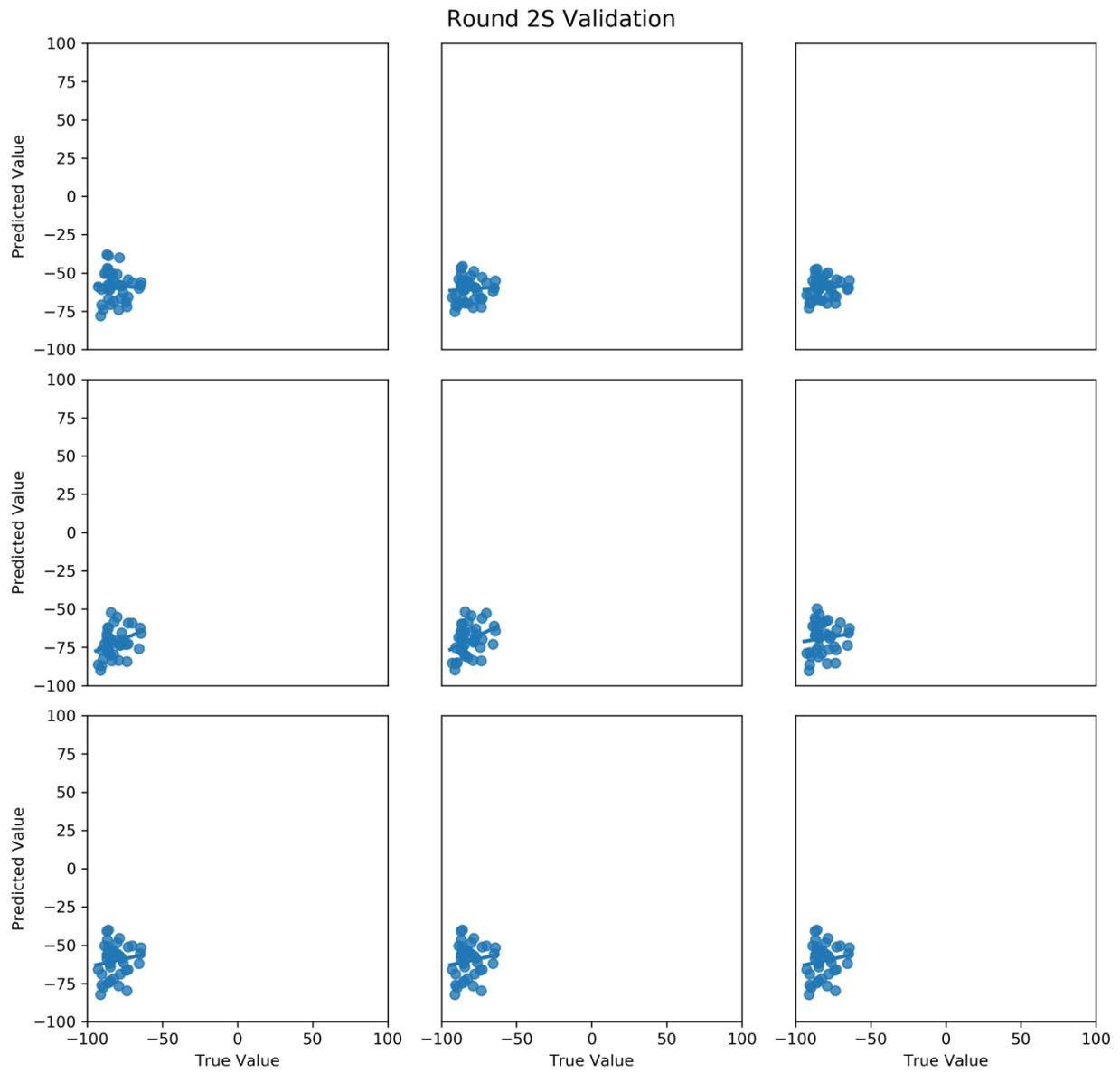

## IV. Input sequences versus encoded predictions

A bias that is not present in the empirical landscape study is in the ratio of sequences that are transformed into the host organism. This ratio can be significantly altered due to cloning biases. Therefore, heat maps of encoded amino acids are shown for each round comparing the input and predicted libraries. (These ratios are often represented with sequence logo maps, which are better visualizations when a few amino acids dominate.)

The input libraries are NDT libraries, which represent N, S, I, H, R, L, D, G, V, Y, C, F. Input libraries also contain proline at position 49 from WT. The degenerate codons used to encode amino acids at each position are provided for reference. Sequence-function data is available on Protabank (8). In these tables, residues in bold were not part of the predicted library but had to be included with the degenerate codon cloning method.

**Figure 3A**: Input versus predicted sequences for modeling position Set I.

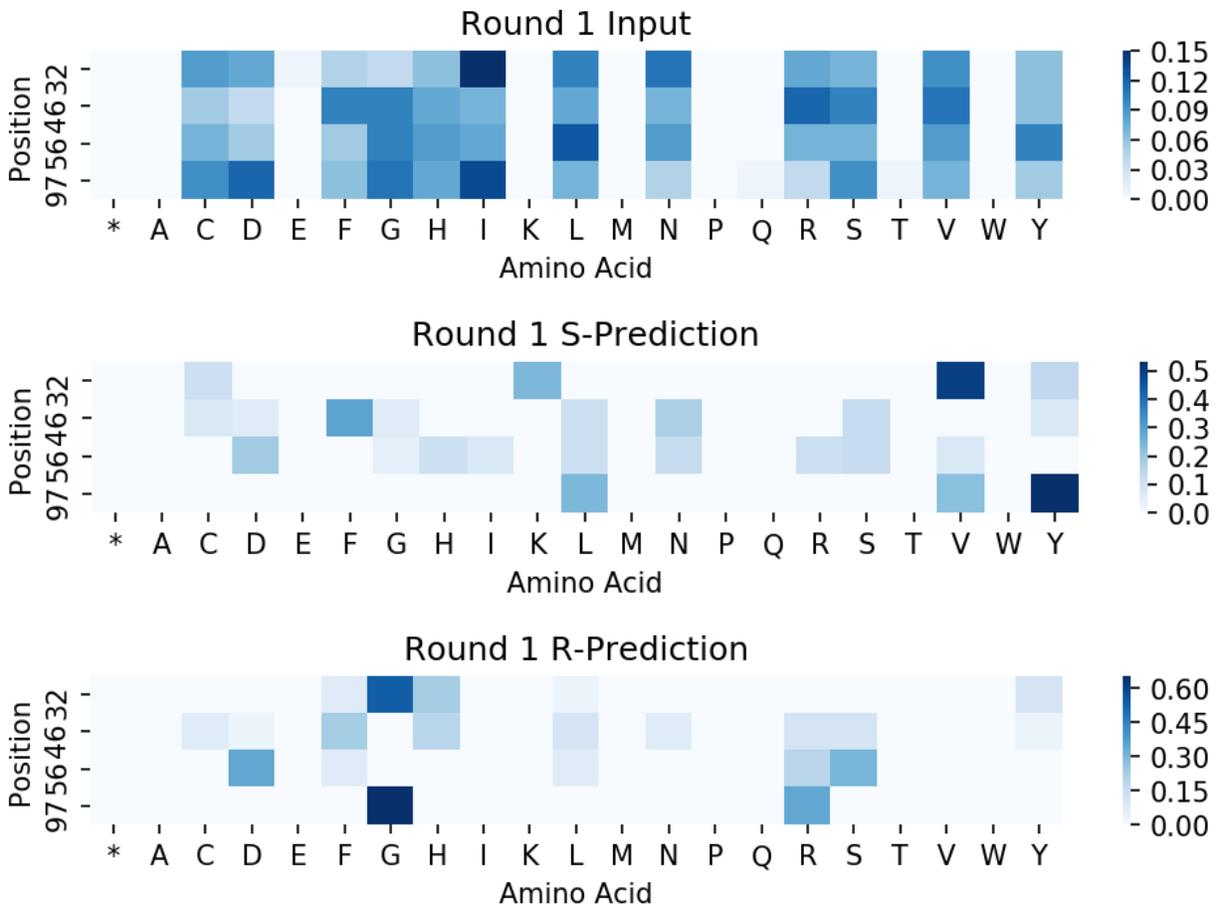

| 1R Predictions | | | 1S Predictions | | |
|---|---|---|---|---|---|
| Position | Codon | Encoded | Position | Codon | Encoded |
| 32 | AAA; GTA; TRC | K; V; C, Y | 32 | GGA; YWC | G; F, H, L, **Y** |
| 46 | DRC; TTM | C, D, G, N, S, Y; F, L | 46 | HDC | C, F, H, I, L, N, **R**, S, Y |
| 56 | VDC | D, G, H, I, L, N, R, S, V | 56 | GAC; YBC | D; C, F, L, **P**, R, S |
| 97 | STA; TAC | L, V; Y | 97 | RGA | G, R |

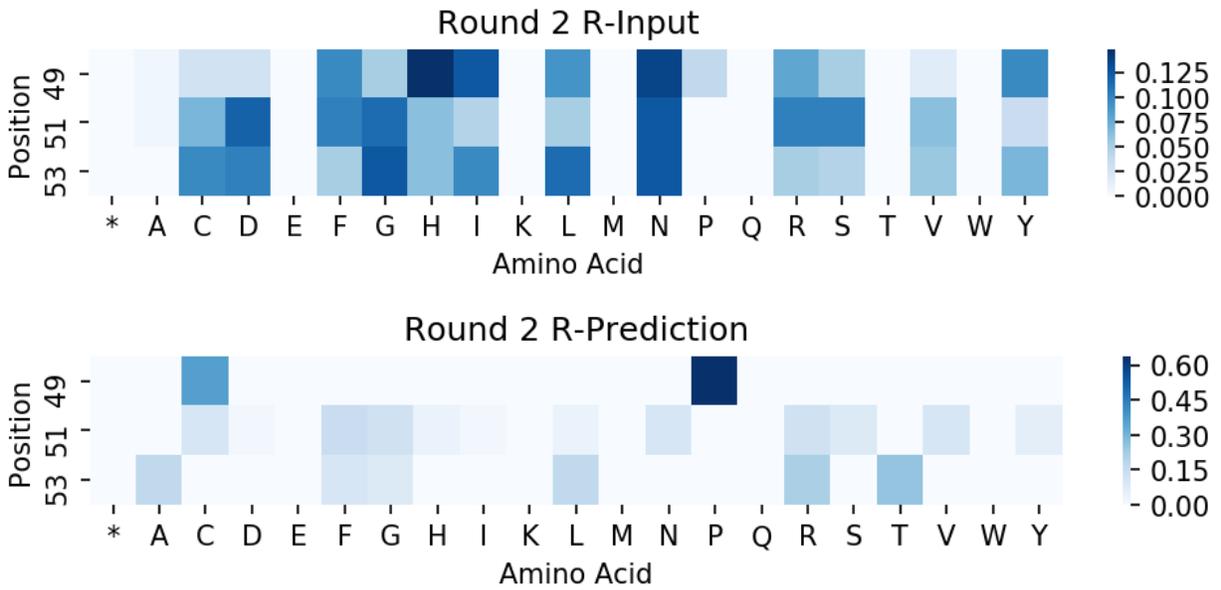

**Figure 3B**: Input versus predicted sequences for modeling Set II from GSSG.

| Position | Codon | Encoded |
|---|---|---|
| 49 | CCA; TGC | P, C |
| 51 | NDT | F, V, Y, N, R, G, I, **H, L, D, C, S** |
| 53 | RSA; TTM | L, F, G, R, **T**; S, C |

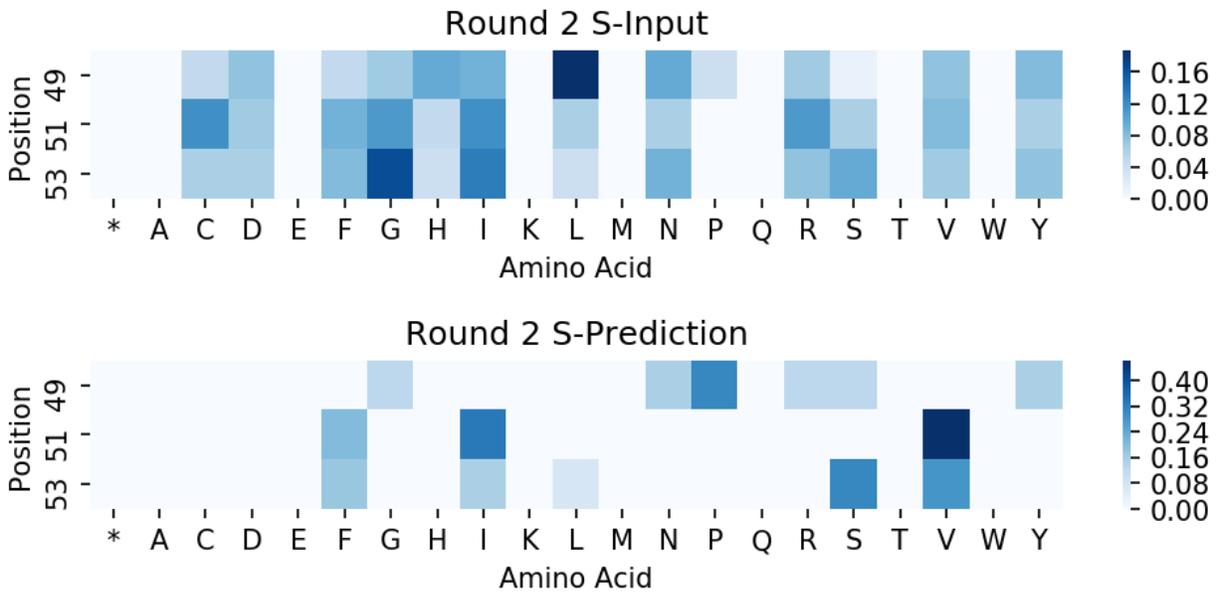

**Figure 3C**: Input versus predicted sequences for modeling Set II from VCHV.

| Position | Codon | Encoded |
|---|---|---|
| 49 | HMC; RGA | Y, N, R; S, P, G |
| 51 | DTC | V, F, I |
| 53 | AGC; NTC | I, V, L; S, F |

## V. Library coverage

The comparison of number of variants necessary for a single-mutation walk is a central argument of the main text and deserves extra explanation. Protein engineers often aim for 95% library coverage (9, 10), or a 95% probability of seeing a particular variant in the library. Assuming equal frequency of each amino acid, this number is roughly 3-fold the library size, which is often used in practice (9). Therefore, for 19 mutations away from one of the 20 canonical amino acids, $19 \times 3 = 57$ variants are needed for roughly 95% coverage. The single-mutation walk to identify mutations at 4 positions has $4 + 3 + 2 + 1 = 10$ such libraries, for a total of 570 variants.

However, a different analysis without making these assumptions can be completed for this particular library by using expressions developed by Bosley and Ostermeier (11). From this work, the probability $P_i$ of a particular sequence $i$ is given below, where $N$ is the number of tested variants and $f_i$ is the frequency at which the sequence $i$ is expected to be present.

$$P_i = 1 - (1 - f_i)^N$$

Rearranged to give

$$N = \frac{\ln(1 - P_i)}{\ln(1 - f_i)}$$

As stated previously, a typical desired library coverage is $P_i = 0.95$ for 95% library coverage, but the choice of codons can have a strong effect on the value of $f_i$. Assuming equal representation of the 19 codons gives $N \approx 55.4$, or 554 variants for the 10 libraries needed. However, the authors of the landscape used NNS/NNK codons, which encode for 20 amino acids with 32 codons. The least frequent amino acid encoded with these codons (methionine) occurs at a frequency of 1/32, requiring $N \approx 94.4$, or 944 variants. A typical balance between balancing the degenerate codon complexity and amino acid coverage that protein engineers employ is the use of NDT/VHG/TGG codons, also known as the 22c-trick (10), in which methionine occurs 1/22 times for $N \approx 64.4$, or 644 variants over 10 libraries.

From a protein engineer's perspective, a comparison to 644 variants is likely the most pertinent in **Supp Figure 1**. However, to provide DE alone with a stronger baseline, we have used 570 variants, obtained from applying the 3-fold oversampling rule (9) to 10 libraries containing 19 desired variants each. as a comparison in the main text. In any case, we empirically observe that the single-mutation walk performs similarly to the ML approach at 300-400 variants (**Supp Figure 1)**, which is significantly less than any of the numbers presented here.

## VI. Chiral SFC traces for racemic and enzymatically synthesized organosilicon products

All the *ee* values of synthesized organosilicon products were determined using automatic peak integration from chiral SFC. The traces for racemic and enzymatic products are shown below. The absolute configuration of products was previously determined (3).

**Chiralcel OD-H (4.6 mm x 25 cm), 5% isopropanol in $CO_2$, 3 mL/min, 210 nm**

### Racemic

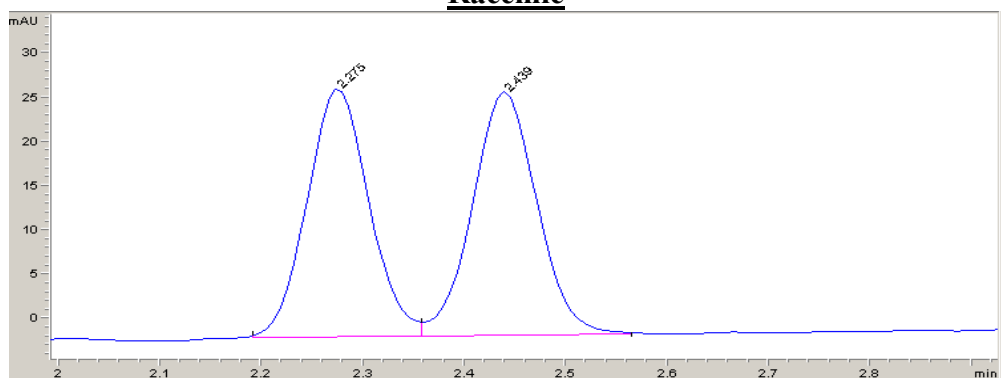

### Variant VCHVYVV

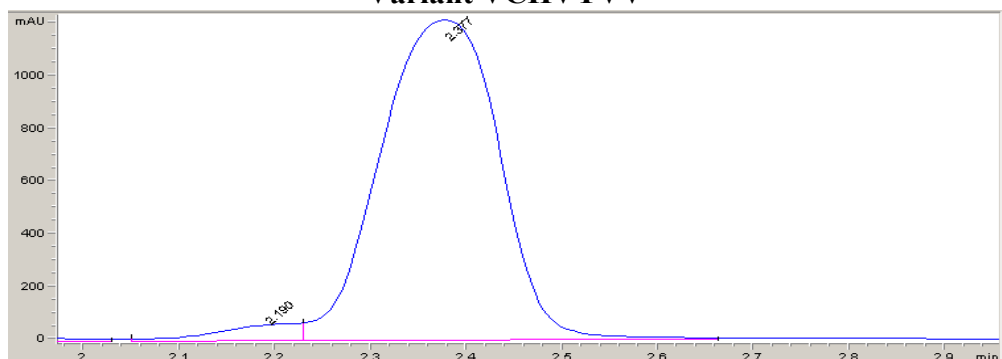

### Variant GSSGPRL

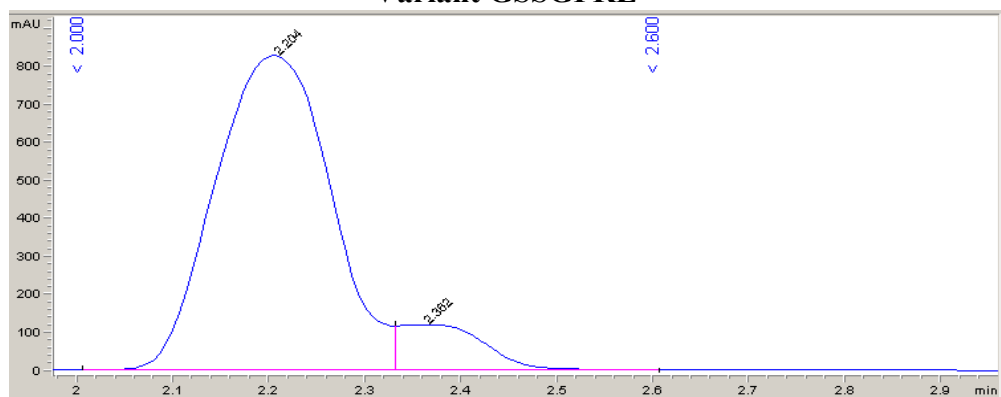

| rac | | | VCHVYVV | | | GSSGPRL | | |
|---|---|---|---|---|---|---|---|---|
| Retention Time (min) | Area (mAU*S) | Area % | Retention Time (min) | Area (mAU*S) | Area % | Retention Time (min) | Area (mAU*S) | Area % |
| 2.275 | 119.2 | 49.4% | 2.19 | 418.3 | 3.8% | 2.204 | 6903.2 | 90.4% |
| 2.439 | 121.9 | 50.6% | 2.377 | 10498.3 | 96.2% | 2.362 | 733.6 | 9.6% |
| Total | 241.1 | | Total | 10916.6 | | Total | 7636.8 | |

## VII. Experimental uncertainty in best *Rma* NOD variants

Although the protein GB1 case study is presented as proof of principle, we also provide evidence that this approach results in significantly improved variants over the input proteins for Position Set II. However, we would like to reiterate that while we have shown this method is more likely to find better variants on an empirical method, this method does not guarantee identifying protein variants that are better than the best identified in the input library. A simple case example is serendipitously identifying the fitness maximum in the input library. The p-values obtained from Welch's *t*-test are shown below.

Activity is significantly improved over starting variant KFLL

| Variant | Mean ± StDev | p-value |
|---------|--------------|---------|
| KFLLPRI | 3290 ± 360   | ---     |
| VCHVYVV | 9330 ± 1780  | 3.77E-02 |
| VCHVPVI | 10670 ± 520  | 2.26E-04 |
| VCHVPVV | 7820 ± 1820  | 6.84E-04 |
| GSSGPRL | 7380 ± 190   | 2.27E-05 |
| GSSGPGL | 7020 ± 230   | 3.37E-05 |
| GSSGPFF | 7300 ± 440   | 6.06E-04 |

Comparisons for enantioselectivity are best done with a different metric than what is typically reported (*ee*). Enantiomeric excess refers to the positive ratio of $|R - S|/(R + S)$. A key assumption of the *t*-test is that each population has a normal distribution, therefore we first convert *ee* to $\Delta\Delta G$ by taking $\ln(R/S)$ where $R$ is the major product or $\ln(S/R)$.

Enantioselectivity in Set II is significantly improved over starting variant VCHV

| Variant | Mean ± StDev of ln(R/S) | p-value |
|---------|--------------------------|---------|
| VCHVPRI | 2.596 ± 0.070            | ---     |
| VCHVYVV | 3.386 ± 0.103            | 1.48E-03 |
| VCHVPVI | 3.213 ± 0.152            | 1.62E-03 |
| VCHVPVV | 3.128 ± 0.010            | 7.54E-03 |

Enantioselectivity in Set II is significantly improved over starting variant GSSG

| Variant | Mean ± StDev of ln(S/R) | p-value |
|---------|--------------------------|---------|
| GSSGPRI | 1.484 ± 0.090            | ---     |
| GSSGPRL | 2.152 ± 0.063            | 1.65E-03 |
| GSSGPGL | 1.925 ± 0.034            | 1.21E-02 |
| GSSGPFF | 1.731 ± 0.062            | 3.93E-02 |

# VIII. Model performance and selection

The LOO Pearson r of the regressors with default hyperparameters and the models ultimately selected are shown for each round.

## Set I Models

| | |
|---|---|
| 0.512212499 | GradientBoostingRegressor(alpha=0.9, criterion='friedman_mse', init=None, learning_rate=0.1, loss='ls', max_depth=3, max_features=None, max_leaf_nodes=None, min_impurity_split=1e-07, min_samples_leaf=1, min_samples_split=2, min_weight_fraction_leaf=0.0, n_estimators=100, presort='auto', random_state=None, subsample=1.0, verbose=0, warm_start=False) |
| 0.478097911 | RandomForestRegressor(bootstrap=True, criterion='mse', max_depth=None, max_features='auto', max_leaf_nodes=None, min_impurity_split=1e-07, min_samples_leaf=1, min_samples_split=2, min_weight_fraction_leaf=0.0, n_estimators=10, n_jobs=1, oob_score=False, random_state=None, verbose=0, warm_start=False) |
| 0.460760125 | LinearSVR(C=1.0, dual=True, epsilon=0.0, fit_intercept=True, intercept_scaling=1.0, loss='epsilon_insensitive', max_iter=1000, random_state=None, tol=0.0001, verbose=0) |
| 0.447166856 | ARDRegression(alpha_1=1e-06, alpha_2=1e-06, compute_score=False, copy_X=True, fit_intercept=True, lambda_1=1e-06, lambda_2=1e-06, n_iter=300, normalize=False, threshold_lambda=10000.0, tol=0.001, verbose=False) |
| 0.423793421 | KernelRidge(alpha=1, coef0=1, degree=3, gamma=None, kernel='linear', kernel_params=None) |
| 0.419172462 | BayesianRidge(alpha_1=1e-06, alpha_2=1e-06, compute_score=False, copy_X=True, fit_intercept=True, lambda_1=1e-06, lambda_2=1e-06, n_iter=300, normalize=False, tol=0.001, verbose=False) |
| 0.406655665 | BaggingRegressor(base_estimator=None, bootstrap=True, bootstrap_features=False, max_features=1.0, max_samples=1.0, n_estimators=10, n_jobs=1, oob_score=False, random_state=None, verbose=0, warm_start=False) |
| 0.396791771 | LassoLarsCV(copy_X=True, cv=None, eps=2.2204460492503131e-16, fit_intercept=True, max_iter=500, max_n_alphas=1000, n_jobs=1, normalize=True, positive=False, precompute='auto', verbose=False) |
| 0.37899373 | DecisionTreeRegressor(criterion='mse', max_depth=None, max_features=None, max_leaf_nodes=None, min_impurity_split=1e-07, min_samples_leaf=1, min_samples_split=2, min_weight_fraction_leaf=0.0, presort=False, random_state=None, splitter='best') |
| 0.371734032 | SGDRegressor(alpha=0.0001, average=False, epsilon=0.1, eta0=0.01, fit_intercept=True, l1_ratio=0.15, learning_rate='invscaling', loss='squared_loss', n_iter=5, penalty='l2', power_t=0.25, random_state=None, shuffle=True, verbose=0, warm_start=False) |
| 0.366256085 | KNeighborsRegressor(algorithm='auto', leaf_size=30, metric='minkowski', metric_params=None, n_jobs=1, n_neighbors=5, p=2, weights='uniform') |
| 0.338423931 | ElasticNet(alpha=1.0, copy_X=True, fit_intercept=True, l1_ratio=0.5, max_iter=1000, normalize=False, positive=False, precompute=False, random_state=None, selection='cyclic', tol=0.0001, warm_start=False) |
| 0.202908183 | AdaBoostRegressor(base_estimator=None, learning_rate=1.0, loss='linear', n_estimators=50, random_state=None) |
| -0.082371549 | LinearRegression(copy_X=True, fit_intercept=True, n_jobs=1, normalize=False) |
| -0.766587492 | NuSVR(C=1.0, cache_size=200, coef0=0.0, degree=3, gamma='auto', kernel='rbf', max_iter=-1, nu=0.5, shrinking=True, tol=0.001, verbose=False) |

## Set I Models Selected

- ARDRegression(alpha_1=0.01, alpha_2=0.1, compute_score=False, copy_X=True, fit_intercept=True, lambda_1=0.01, lambda_2=0.01, n_iter=300, normalize=False, threshold_lambda=10000.0, tol=0.0001, verbose=False)
- ARDRegression(alpha_1=0.01, alpha_2=0.01, compute_score=False, copy_X=True, fit_intercept=True, lambda_1=0.01, lambda_2=0.01, n_iter=300, normalize=False, threshold_lambda=10000.0, tol=0.0001, verbose=False)
- ARDRegression(alpha_1=0.01, alpha_2=0.0001, compute_score=False, copy_X=True, fit_intercept=True, lambda_1=0.01, lambda_2=0.01, n_iter=300, normalize=False, threshold_lambda=10000.0, tol=0.0001, verbose=False)
- GradientBoostingRegressor(alpha=0.3, criterion='mse', init=None,learning_rate=0.9, loss='quantile', max_depth=3,max_features=None, max_leaf_nodes=None,min_impurity_split=1e-07, min_samples_leaf=1,min_samples_split=2, min_weight_fraction_leaf=0.0,n_estimators=500, presort='auto', random_state=None, subsample=1.0, verbose=0, warm_start=False)
- GradientBoostingRegressor(alpha=0.5, criterion='mse', init=None,learning_rate=0.7, loss='huber', max_depth=10,max_features=None, max_leaf_nodes=None,min_impurity_split=1e-07, min_samples_leaf=1,min_samples_split=2, min_weight_fraction_leaf=0.0,n_estimators=1000, presort='auto', random_state=None,subsample=1.0, verbose=0, warm_start=False)
- GradientBoostingRegressor(alpha=0.5, criterion='mse', init=None,learning_rate=0.7, loss='huber', max_depth=10,max_features=None, max_leaf_nodes=None,min_impurity_split=1e-07, min_samples_leaf=1,min_samples_split=2, min_weight_fraction_leaf=0.0,n_estimators=100, presort='auto', random_state=None,subsample=1.0, verbose=0, warm_start=False)
- LinearSVR(C=50, dual=True, epsilon=0, fit_intercept=True,intercept_scaling=1.0, loss='epsilon_insensitive', max_iter=10000,random_state=None, tol=0.0001, verbose=0)
- LinearSVR(C=50, dual=True, epsilon=0.1, fit_intercept=True,intercept_scaling=1.0, loss='epsilon_insensitive', max_iter=10000,random_state=None, tol=0.0001, verbose=0)
- LinearSVR(C=1.0, dual=True, epsilon=0.0, fit_intercept=True,intercept_scaling=1.0, loss='epsilon_insensitive', max_iter=1000,random_state=None, tol=0.0001, verbose=0)

## Set 2(*S*) Models

| | |
|---|---|
| 0.609011 | ARDRegression(alpha_1=1e-06, alpha_2=1e-06, compute_score=False, copy_X=True, fit_intercept=True, lambda_1=1e-06, lambda_2=1e-06, n_iter=300, normalize=False, threshold_lambda=10000.0, tol=0.001, verbose=False) |
| 0.60823 | NuSVR(C=1.0, cache_size=200, coef0=0.0, degree=3, gamma='auto', kernel='rbf', max_iter=-1, nu=0.5, shrinking=True, tol=0.001, verbose=False) |
| 0.598652 | LinearSVR(C=1.0, dual=True, epsilon=0.0, fit_intercept=True, intercept_scaling=1.0, loss='epsilon_insensitive', max_iter=1000, random_state=None, tol=0.0001, verbose=0) |
| 0.587369 | KernelRidge(alpha=1, coef0=1, degree=3, gamma=None, kernel='linear', kernel_params=None) |
| 0.584004 | BayesianRidge(alpha_1=1e-06, alpha_2=1e-06, compute_score=False, copy_X=True, fit_intercept=True, lambda_1=1e-06, lambda_2=1e-06, n_iter=300, normalize=False, tol=0.001, verbose=False) |
| 0.578776 | GradientBoostingRegressor(alpha=0.9, criterion='friedman_mse', init=None, learning_rate=0.1, loss='ls', max_depth=3, max_features=None, max_leaf_nodes=None, min_impurity_decrease=0.0, min_impurity_split=None, min_samples_leaf=1, min_samples_split=2, min_weight_fraction_leaf=0.0, n_estimators=100, presort='auto', random_state=None, subsample=1.0, verbose=0, warm_start=False) |
| 0.577483 | LinearRegression(copy_X=True, fit_intercept=True, n_jobs=1, normalize=False) |
| 0.564136 | BaggingRegressor(base_estimator=None, bootstrap=True, bootstrap_features=False, max_features=1.0, max_samples=1.0, n_estimators=10, n_jobs=1, oob_score=False, random_state=None, verbose=0, warm_start=False) |
| 0.549607 | RandomForestRegressor(bootstrap=True, criterion='mse', max_depth=None, max_features='auto', max_leaf_nodes=None, min_impurity_decrease=0.0, min_impurity_split=None, min_samples_leaf=1, min_samples_split=2, min_weight_fraction_leaf=0.0, n_estimators=10, n_jobs=1, oob_score=False, random_state=None, verbose=0, warm_start=False) |
| 0.503091 | MLPRegressor(activation='relu', alpha=0.0001, batch_size='auto', beta_1=0.9, beta_2=0.999, early_stopping=False, epsilon=1e-08, hidden_layer_sizes=(100,), learning_rate='constant', learning_rate_init=0.001, max_iter=200, momentum=0.9, nesterovs_momentum=True, power_t=0.5, random_state=None, shuffle=True, solver='adam', tol=0.0001, validation_fraction=0.1, verbose=False, warm_start=False) |
| 0.499812 | DecisionTreeRegressor(criterion='mse', max_depth=None, max_features=None, max_leaf_nodes=None, min_impurity_decrease=0.0, min_impurity_split=None, min_samples_leaf=1, min_samples_split=2, min_weight_fraction_leaf=0.0, presort=False, random_state=None, splitter='best') |
| 0.438121 | LassoLarsCV(copy_X=True, cv=None, eps=2.2204460492503131e-16, fit_intercept=True, max_iter=500, max_n_alphas=1000, n_jobs=1, normalize=True, positive=False, precompute='auto', verbose=False) |
| 0.437293 | SGDRegressor(alpha=0.0001, average=False, epsilon=0.1, eta0=0.01, fit_intercept=True, l1_ratio=0.15, learning_rate='invscaling', loss='squared_loss', max_iter=None, n_iter=None, penalty='l2', power_t=0.25, random_state=None, shuffle=True, tol=None, verbose=0, warm_start=False) |
| 0.430372 | KNeighborsRegressor(algorithm='auto', leaf_size=30, metric='minkowski', metric_params=None, n_jobs=1, n_neighbors=5, p=2, weights='uniform') |

| 0.399228 | AdaBoostRegressor(base_estimator=None, learning_rate=1.0, loss='linear', n_estimators=50, random_state=None) |
|---|---|

## Set 2(*S*) Models Selected

- GradientBoostingRegressor(alpha=0.1, criterion='mse', init=None, learning_rate=0.7, loss='lad', max_depth=3, max_features=None, max_leaf_nodes=None, min_impurity_decrease=0.0, min_impurity_split=None, min_samples_leaf=1, min_samples_split=2, min_weight_fraction_leaf=0.0, n_estimators=100, presort='auto', random_state=None, subsample=1.0, verbose=0, warm_start=False)
- GradientBoostingRegressor(alpha=0.5, criterion='friedman_mse', init=None, learning_rate=0.9, loss='quantile', max_depth=3, max_features=None, max_leaf_nodes=None, min_impurity_decrease=0.0, min_impurity_split=None, min_samples_leaf=1, min_samples_split=2, min_weight_fraction_leaf=0.0, n_estimators=100, presort='auto', random_state=None, subsample=1.0, verbose=0, warm_start=False)
- GradientBoostingRegressor(alpha=0.3, criterion='friedman_mse', init=None, learning_rate=0.3, loss='quantile', max_depth=3, max_features=None, max_leaf_nodes=None, min_impurity_decrease=0.0, min_impurity_split=None, min_samples_leaf=1, min_samples_split=2, min_weight_fraction_leaf=0.0, n_estimators=100, presort='auto', random_state=None, subsample=1.0, verbose=0, warm_start=False)
- ARDRegression(alpha_1=0.1, alpha_2=1e-08, compute_score=False, copy_X=True, fit_intercept=True, lambda_1=0.1, lambda_2=1e-06, n_iter=3000, normalize=False, threshold_lambda=10000.0, tol=0.0001, verbose=False)
- ARDRegression(alpha_1=0.1, alpha_2=1e-08, compute_score=False, copy_X=True, fit_intercept=True, lambda_1=0.1, lambda_2=1e-06, n_iter=300, normalize=False, threshold_lambda=10000.0, tol=0.0001, verbose=False)
- ARDRegression(alpha_1=0.1, alpha_2=1e-06, compute_score=False, copy_X=True, fit_intercept=True, lambda_1=0.1, lambda_2=1e-06, n_iter=3000, normalize=False, threshold_lambda=10000.0, tol=0.0001, verbose=False)
- LinearSVR(C=50, dual=True, epsilon=0, fit_intercept=True, intercept_scaling=1.0, loss='epsilon_insensitive', max_iter=10000, random_state=None, tol=0.0001, verbose=0)
- LinearSVR(C=100, dual=True, epsilon=0, fit_intercept=True, intercept_scaling=1.0, loss='epsilon_insensitive', max_iter=10000, random_state=None, tol=0.0001, verbose=0)
- LinearSVR(C=1000, dual=True, epsilon=0, fit_intercept=True, intercept_scaling=1.0, loss='squared_epsilon_insensitive', max_iter=10000, random_state=None, tol=0.0001, verbose=0)

**Set 2(*R*) Models**

| | |
|---|---|
| 0.651306 | GradientBoostingRegressor(alpha=0.9, criterion='friedman_mse', init=None, learning_rate=0.1, loss='ls', max_depth=3, max_features=None, max_leaf_nodes=None, min_impurity_decrease=0.0, min_impurity_split=None, min_samples_leaf=1, min_samples_split=2, min_weight_fraction_leaf=0.0, n_estimators=100, presort='auto', random_state=None, subsample=1.0, verbose=0, warm_start=False) |
| 0.635081 | ARDRegression(alpha_1=1e-06, alpha_2=1e-06, compute_score=False, copy_X=True, fit_intercept=True, lambda_1=1e-06, lambda_2=1e-06, n_iter=300, normalize=False, threshold_lambda=10000.0, tol=0.001, verbose=False) |
| 0.631197 | BayesianRidge(alpha_1=1e-06, alpha_2=1e-06, compute_score=False, copy_X=True, fit_intercept=True, lambda_1=1e-06, lambda_2=1e-06, n_iter=300, normalize=False, tol=0.001, verbose=False) |
| 0.626982 | KernelRidge(alpha=1, coef0=1, degree=3, gamma=None, kernel='linear', kernel_params=None) |
| 0.625465 | AdaBoostRegressor(base_estimator=None, learning_rate=1.0, loss='linear', n_estimators=50, random_state=None) |
| 0.61362 | NuSVR(C=1.0, cache_size=200, coef0=0.0, degree=3, gamma='auto', kernel='rbf', max_iter=-1, nu=0.5, shrinking=True, tol=0.001, verbose=False) |
| 0.612285 | LassoLarsCV(copy_X=True, cv=None, eps=2.2204460492503131e-16, fit_intercept=True, max_iter=500, max_n_alphas=1000, n_jobs=1, normalize=True, positive=False, precompute='auto', verbose=False) |
| 0.608155 | RandomForestRegressor(bootstrap=True, criterion='mse', max_depth=None, max_features='auto', max_leaf_nodes=None, min_impurity_decrease=0.0, min_impurity_split=None, min_samples_leaf=1, min_samples_split=2, min_weight_fraction_leaf=0.0, n_estimators=10, n_jobs=1, oob_score=False, random_state=None, verbose=0, warm_start=False) |
| 0.595166 | MLPRegressor(activation='relu', alpha=0.0001, batch_size='auto', beta_1=0.9, beta_2=0.999, early_stopping=False, epsilon=1e-08, hidden_layer_sizes=(100,), learning_rate='constant', learning_rate_init=0.001, max_iter=200, momentum=0.9, nesterovs_momentum=True, power_t=0.5, random_state=None, shuffle=True, solver='adam', tol=0.0001, validation_fraction=0.1, verbose=False, warm_start=False) |
| 0.583177 | BaggingRegressor(base_estimator=None, bootstrap=True, bootstrap_features=False, max_features=1.0, max_samples=1.0, n_estimators=10, n_jobs=1, oob_score=False, random_state=None, verbose=0, warm_start=False) |
| 0.552808 | LinearSVR(C=1.0, dual=True, epsilon=0.0, fit_intercept=True, intercept_scaling=1.0, loss='epsilon_insensitive', max_iter=1000, random_state=None, tol=0.0001, verbose=0) |
| 0.542432 | SGDRegressor(alpha=0.0001, average=False, epsilon=0.1, eta0=0.01, fit_intercept=True, l1_ratio=0.15, learning_rate='invscaling', loss='squared_loss', max_iter=None, n_iter=None, penalty='l2', power_t=0.25, random_state=None, shuffle=True, tol=None, verbose=0, warm_start=False) |
| 0.479498 | DecisionTreeRegressor(criterion='mse', max_depth=None, max_features=None, max_leaf_nodes=None, min_impurity_decrease=0.0, min_impurity_split=None, min_samples_leaf=1, min_samples_split=2, min_weight_fraction_leaf=0.0, presort=False, random_state=None, splitter='best') |
| 0.473718 | KNeighborsRegressor(algorithm='auto', leaf_size=30, metric='minkowski', metric_params=None, n_jobs=1, n_neighbors=5, p=2, weights='uniform') |
| 0.242713 | LinearRegression(copy_X=True, fit_intercept=True, n_jobs=1, normalize=False) |

**Set 2(*R*) Models Selected**

- GradientBoostingRegressor(alpha=0.1, criterion='friedman_mse', init=None, learning_rate=0.1, loss='ls', max_depth=3, max_features=None, max_leaf_nodes=None, min_impurity_decrease=0.0, min_impurity_split=None, min_samples_leaf=1, min_samples_split=2, min_weight_fraction_leaf=0.0, n_estimators=100, presort='auto', random_state=None, subsample=1.0, verbose=0, warm_start=False)
- GradientBoostingRegressor(alpha=0.5, criterion='mse', init=None, learning_rate=0.1, loss='ls', max_depth=3, max_features=None, max_leaf_nodes=None, min_impurity_decrease=0.0, min_impurity_split=None, min_samples_leaf=1, min_samples_split=2, min_weight_fraction_leaf=0.0, n_estimators=500, presort='auto', random_state=None, subsample=1.0, verbose=0, warm_start=False)
- GradientBoostingRegressor(alpha=0.7, criterion='friedman_mse', init=None, learning_rate=0.1, loss='ls', max_depth=3, max_features=None, max_leaf_nodes=None, min_impurity_decrease=0.0, min_impurity_split=None, min_samples_leaf=1, min_samples_split=2, min_weight_fraction_leaf=0.0, n_estimators=100, presort='auto', random_state=None, subsample=1.0, verbose=0, warm_start=False)
- ARDRegression(alpha_1=1, alpha_2=1e-08, compute_score=False, copy_X=True, fit_intercept=True, lambda_1=1e-08, lambda_2=0.1, n_iter=10000, normalize=False, threshold_lambda=10000.0, tol=0.0001, verbose=False)
- ARDRegression(alpha_1=1, alpha_2=0.0001, compute_score=False, copy_X=True, fit_intercept=True, lambda_1=1e-08, lambda_2=0.1, n_iter=10000, normalize=False, threshold_lambda=10000.0, tol=0.0001, verbose=False)
- ARDRegression(alpha_1=1, alpha_2=1e-08, compute_score=False, copy_X=True, fit_intercept=True, lambda_1=0.0001, lambda_2=0.1, n_iter=10000, normalize=False, threshold_lambda=10000.0, tol=0.0001, verbose=False)
- BayesianRidge(alpha_1=0.1, alpha_2=1e-08, compute_score=False, copy_X=True, fit_intercept=True, lambda_1=1e-08, lambda_2=0.1, n_iter=10000, normalize=False, tol=0.0001, verbose=False)
- BayesianRidge(alpha_1=0.1, alpha_2=1e-08, compute_score=False, copy_X=True, fit_intercept=True, lambda_1=1e-08, lambda_2=0.1, n_iter=3000, normalize=False, tol=0.0001, verbose=False)
- BayesianRidge(alpha_1=0.1, alpha_2=1e-08, compute_score=False, copy_X=True, fit_intercept=True, lambda_1=1e-08, lambda_2=0.1, n_iter=300, normalize=False, tol=0.0001, verbose=False)